\documentclass[a4paper]{article}

\usepackage{microtype}
\usepackage[sc]{mathpazo}
\usepackage[T1]{fontenc}
\usepackage[DIV=11,BCOR=0cm,headinclude=true]{typearea}

\usepackage{amsfonts}
\usepackage{amsmath}
\usepackage{amssymb}
\usepackage{amsthm}
\usepackage{graphicx}
\usepackage{float}
\usepackage[config,textfont=sl]{caption,subfig}
\usepackage[british]{babel}
\usepackage{bm}

\usepackage[maxbibnames=100,autocite=superscript,style=custom-nature,sorting=none,defernumbers=true]{biblatex}

\usepackage{authblk}

\newcommand{\bv}[1]{\boldsymbol{#1}}
\newcommand{\bra}[1]{\langle{#1}|}
\newcommand{\ket}[1]{|{#1}\rangle}
\newcommand{\braket}[2]{\langle{#1}|{#2}\rangle}
\newcommand{\tensor}[1]{\bar{\bar{#1}}}

\DeclareMathOperator*{\real}{Re}
\DeclareMathOperator*{\imag}{Im}

\author[1,$\dagger$]{Parry Y.\ Chen}
\author[1]{Yonatan Sivan}
\author[2]{Egor A.\ Muljarov}
\affil[1]{School of Electrical and Computer Engineering, Ben-Gurion University, Israel}
\affil[2]{School of Physics and Astronomy, Cardiff University, Cardiff, United Kingdom}
\affil[$\dagger$]{\it{parryyu@post.bgu.ac.il}}

\def\XXint#1#2#3{{\setbox0=\hbox{$#1{#2#3}{\int}$}
     \vcenter{\hbox{$#2#3$}}\kern-.5\wd0}}

\title{An efficient solver for the generalized normal modes of non-uniform open optical resonators}
\date{\today}

\begin{document}
\maketitle

\begin{abstract}
Modal expansion is an attractive technique for solving electromagnetic scattering problems. With the one set of resonator modes, calculated once and for all, any configuration of near-field or far-field sources can be obtained almost instantaneously. Traditionally applied to closed systems, a simple and rigorous generalization of modal expansion to open systems using eigenpermittivity states is also available. These open modes are suitable for typical nanophotonic systems, for example. However, the numerical generation of modes is usually the most difficult and time-consuming step of modal expansion techniques. Here, we demonstrate efficient and reliable mode generation, expanding the target modes into the modes of a simpler open system that are known. Such a re-expansion technique is implemented for resonators with non-uniform permittivity profiles, demonstrating its rapid convergence. Key to the method's success is the inclusion of a set of longitudinal basis modes.
\end{abstract}

\section{Introduction}
Modal expansion methods have long been used to solve partial differential equations (PDEs) with source terms,\autocite{morse1946methods} and are attractive alternatives to direct solutions of PDEs with sources. Thus, instead of solving a scattering problem with a specific source term, first the eigennatural modes of the governing PDE are found by setting source terms to zero, yielding an eigenvalue problem. The resulting eigenmodes represent the resonances supported by the system in the absence of any external excitation. The eigenmodes then serve as the perfect basis for representing the solution of the scattering problem for any possible configuration of sources.

Having found the eigenmodes of the system, no further numerical computation is required, even as the source position and orientation are changed.\autocite{muljarov2010brillouin, doost2013resonant, lalanne2018light, lalanne2018rigorous} Since the eigenmodes are calculated once and for all, the initial time invested in finding the modes is often rewarded with significant time savings, greatly expediting computationally intensive simulations. In optics, this includes for example Green's tensor calculations for quantum optical effects on the nanoscale,\autocite{novotny2012principles,muljarov2016exact} optimization of optical systems, and inverse problems.\autocite{molesky2018inverse} Particularly in resonant nanophotonic systems, only a few modes are required to capture most of the response, leading to great physical insight.

Modal expansions are usually easy to use once the eigenmodes are available, but finding the modes is often the most challenging and time-consuming step. One popular option is to descretize space into small domains, giving rise to finite element methods (FEMs),\autocite{jin2015finite} for example. For Maxwell's equations, many popular commercial software packages exist, such as COMSOL Multiphysics and CST Microwave Studio. These convert Maxwell's equations to a large but sparse linear eigenvalue problem, $\tensor{A}\bv{x}_m = \lambda_m\bv{x}_m$. However, the numerical algorithms for eigenvalue problems can be slower by two to three orders of magnitude compared to the equivalent direct problem, $\bv{x} = \tensor{A}^{-1} \bv{b}$, negating much of the speed advantage of modal expansion methods. They are also usually memory intensive, necessitating the use of large computing clusters, particularly for challenging problems such as multi-scale geometries.\autocite{solis2014toward} 

A more serious issue is reliability. Large sparse eigenproblems are usually solved by iterative algorithms that require an initial guess,\autocite{trefethen1997numerical} so it is difficult to guarantee that all modes necessary for expansion have been found. Furthermore, numerical noise causes many unphysical or spurious modes to be obtained. These are either artificially localized or longitudinal and their presence pollutes the modal sum. They can be difficult to discern from true modes, yet still need to be identified and discarded, a process that often requires manual inspection and experienced judgment.\autocite{demesy2018nonlinear} These problems are particularly pronounced when searching for plasmonic modes. When high accuracy is desired, the algorithm often fails to converge altogether, precluding use of the modal approach. The lack of fast, reliable, and general numerical schemes for obtaining eigenmodes impedes the widespread use of modal methods.

Another difficulty of modal expansion methods arises when treating open systems, typically consisting of a finite resonator or inclusion in an otherwise infinite background. Such configurations are particularly topical for nanophotonics. Over the preceding few years, modal expansions for open systems have attracted intensive research effort, and significant progress has been achieved. Since the resonator is constantly leaking energy to the background, one approach is to use complex eigenfrequency modes to account for this lack of energy conservation.\autocite{lalanne2018rigorous, muljarov2016resonant, ge2016quasinormal, zschiedrich2018riesz} The imaginary part of the eigenfrequency relates to the finite lifetime or, conversely, the decay rate of the mode. These methods were introduced by Gamow and Siegert in the context of nuclear physics,\autocite{gamow1928zur,siegert1939derivation} and were developed further in quantum mechanics\autocite{baz1969scattering} and wave optics.\autocite{weinstein1969open}

However, expansion by complex frequency modes provides an accurate solution only inside the resonator interior. In the case of material dispersion, the modes are also defined by a nonlinear eigenvalue problem, requiring use of sum rules and alternative expansions of Green's tensor,\autocite{muljarov2016resonant} or auxiliary fields,\autocite{lalanne2019quasinormal} to linearize. Moreover, exponentially large errors develop far from the scatterers, including regimes where the field is typically measured experimentally \autocite{lalanne2018light, lalanne2018rigorous}. One way to circumvent this problem was proposed by Yan et al.\autocite{yan2018rigorous} It involves enclosing the simulation domain with a perfectly matched layer (PML), and including modes that reside primarily within the PML as part of the modal expansion. Unfortunately, such modes are devoid of any physical meaning, and a large number of them is needed, diluting the physical insight that modal expansions typically provide.

We have recently demonstrated that a slightly different modal method bypasses these difficulties of complex frequency modes. We reformulate the eigenvalue problem in terms of optical properties of the scatterer, specifying the refractive index or permittivity to be the eigenvalue.\autocite{bergman1992solid, agranovich1999generalized, chen2019generalizing, sandu2012eigenmode, ge2010steady} This eigenpermittivity is in general complex, but is associated only with the finite interior of the scatterer. All other quantities remain real, such as the frequency, thus defining true normal modes for open systems that neither decay in time nor diverge in space. Furthermore, the propagation constant of each mode within the infinite background matches that of the target problem, which not only avoids the continuum of modes usually associated with infinite space, but allows the set of modes to remain discrete. The fields are correctly reproduced everywhere in space, and also the source term can be placed anywhere in space.

In this sense, our formulation provides a generalization of normal modes to open systems: one that jettisons frequency as the eigenvalue, but in return retains the simplicity and rigor enjoyed by normal modes within closed systems. Thus, we entitled our method Generalized Normal Mode Expansion (GENOME). Its range of practical and theoretical advantages has previously been discussed.\autocite{chen2019generalizing} Finally, we mention that since permittivity in optics is analogous to potential in quantum mechanics, equivalent normal modes exist for open quantum systems. All variables associated with the target problem would be held constant except for the potential, which would be multiplied by a complex factor until the system comes to resonance.

The method was successfully applied to a series of simple cases,\autocite{agranovich1999generalized} including 1D slabs,\autocite{farhi2016electromagnetic} 2D wires,\autocite{chen2019generalizing,reddy2017revisiting} and spheres,\autocite{bergman1980theory, forestiere2016material, pascale2019full} where the modes could be found easily and quickly via a transcendental equation, for which we built a reliable root search algorithm.\autocite{chen2017robust} For modes of a general geometry we adapted the eigenfrequency solvers of COMSOL, a FEM-based software package, to produce eigenpermittivity modes.\autocite{chen2019generalizing, rosolen2019overcoming} Despite the rapid implementation and ease of use, the COMSOL implementation nevertheless suffered from many of the aforementioned issues. 

\begin{figure}[!t]
	\begin{center}
		\includegraphics[width=80mm]{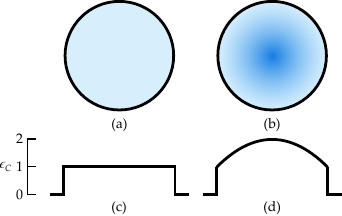}
		\caption{Basis modes of a structure with a uniform permittivity profile (a) are used to expand target modes of a structure with a spatially varying permittivity profile (b). The exterior surfaces of both structures are identical, and both rest in an infinite uniform background. Shown in (c) and (d) are cross sections of the permittivity profiles along their respective diameters. Permittivity contrasts are normalized by background permittivities, $\epsilon_C = \epsilon/\epsilon_b-1$.}
		\label{fig:profile}
	\end{center}
\end{figure}

In this paper, we aim to construct an alternative method for generating eigenmodes that is fast, accurate, reliable, and general, thus overcoming the barriers to widespread adoption of GENOME. We present a method generating the modes of a resonator whose interior exhibits a smoothly varying permittivity profile. A step index change can exist along its exterior surface (e.g. the geometry of Figure \ref{fig:profile} (b) and (d)). This includes as a subset resonators with completely smooth permittivity profiles that are compactly supported.

In this paper, we shall undertake two tasks. Firstly, we generalize GENOME to handle resonators with spatially varying permittivity profiles. The formalism changes only minimally compared to Ref.\ \parencite{chen2019generalizing}, but requires a new set of modes specific to spatially varying profiles. Secondly, we develop a numerical method to find the appropriate eigenmodes, which shall occupy the majority of the paper. This method is based on the concept of resonant state expansion by E.\ A.\ Muljarov, W.\ Langbein, and R.\ Zimmermann, developed in the context of complex frequency modes.\autocite{muljarov2010brillouin}

We obtain modes of the target structure by expanding in terms of a set of basis modes of a simpler open system,\autocite{bergman1980theory, muljarov2010brillouin}
\begin{equation}
	\bv{E}_m(\bv{r}) = \sum_\mu c_{\mu,m} \tilde{\bv{E}}_\mu(\bv{r}).
	\label{eq:expansion}
\end{equation}
We shall use the modes of a resonator with a uniform interior, $\{\tilde{\bv{E}}_\mu(\bv{r})\}$, as a basis to construct the modes of a resonator with the same exterior surface, but with a smoothly varying interior permittivity profile, $\bv{E}_m(\bv{r})$. These are depicted in Figure \ref{fig:profile}. We dub the process \emph{re-expansion}, since we are using basis modes to expand target modes, which are in turn used to expand solutions to Maxwell's equations.

The re-expansion method is a series expansion solution to an eigenvalue problem, and bears many similarities to other methods such as Fourier series expansion. One distinguishing feature is our use of the modes of a simpler open system as a basis, which has several important advantages over the more familiar Fourier basis. These basis modes form a complete yet discrete set, with each of them already satisfying the Sommerfeld radiation condition at infinity. The foregoing also means that the re-expansion method resembles quantum mechanical perturbation theory,\autocite{cohen1977quantum} and its implementation in electrodynamics,\autocite{koegelnik1969coupled, yariv2007photonics, haus1984waves, chuang1987coupled, hardy1987reformulation, haus1989coupled, kocabas2009modal, okamoto2006fundamentals, snyder1983optical} though it differs in some key respects. Firstly, perturbation methods are typically used for weak perturbations, where only the first few terms of the perturbation series are retained. Our method is capable of treating perturbations of any strength, often obtaining rapid convergence towards the true solution. Indeed, our method is equivalent to perturbation to all orders, as it involves a matrix diagonalization. Secondly, we apply it to open systems, whereas standard perturbation methods treat only closed systems associated with Hermitian operators.

The paper is organized as follows. In Section \ref{sec:smooth}, we generalize GENOME to treat resonators with spatially varying permittivity profiles. In Section \ref{sec:perturbation}, we develop the re-expansion method to find the generalized normal modes. In Section \ref{sec:basis}, we detail the basis modes required as inputs to the re-expansion method. In Section \ref{sec:numerics}, we demonstrate the implementation of Sections \ref{sec:smooth}--\ref{sec:basis} and provide numerical examples and convergence data. Finally, we summarize and conclude in Section \ref{sec:conc}.

\section{Generalized normal mode expansion for spatially varying resonators}
\label{sec:smooth}
Our goal is to solve Maxwell's equations with an arbitrary source $\bv{J}(\bv{r})$,
\begin{equation}
\nabla\times(\nabla\times\bv{E}) - k^2\epsilon(\bv{r})\bv{E} = ikZ_0\bv{J},
\label{eq:maxwell}
\end{equation}
where $k = \omega/c$ and $Z_0 = \sqrt{\mu_0/\epsilon_0}$ is the impedance of free space in SI units. We have assumed harmonic $e^{-i\omega t}$ time variation, and also assume non-magnetic media across the whole domain. The geometry is defined by its permittivity profile $\epsilon(\bv{r})$, which can vary arbitrarily subject only to the restrictions that the resonator be finite in extent so that the normalized permittivity contrast
\begin{equation}
\epsilon_C(\bv{r}) = \frac{\epsilon(\bv{r})-\epsilon_b}{\epsilon_b}
\label{eq:epsC}
\end{equation}
is a compactly supported function.

The formulation generalizes previous derivations.\autocite{chen2019generalizing} It first expresses \eqref{eq:maxwell} as a Lippmann-Schwinger equation, then expanding the solution in terms of its eigenmodes. We thus manipulate $\eqref{eq:maxwell}$ to yield
\begin{equation}
\nabla\times(\nabla\times\bv{E}) - k^2\epsilon_b\bv{E} = ikZ_0\bv{J} + k^2\epsilon_b\epsilon_C(\bv{r})\bv{E},
\label{eq:inhomowave}
\end{equation}
where $\epsilon_C(\bv{r})$ is the normalized permittivity contrast. Since the operator on the LHS of \eqref{eq:inhomowave} is now uniform, solution via the Green's function for uniform media is possible,
\begin{equation}
\nabla\times(\nabla\times\tensor{G}_0) - k^2\epsilon_b\tensor{G}_0 = \tensor{I}\delta^3(\bv{r}-\bv{r}'),
\label{eq:greensfree}
\end{equation}
which has a simple known analytic form $\tensor{G}_0(|\bv{r}-\bv{r}'|)$ depending on the dimensionality of the problem. Applying \eqref{eq:greensfree} to both terms on the RHS of \eqref{eq:inhomowave} yields its Green's function solution, the desired Lippmann-Schwinger equation, 
\begin{equation}
\bv{E}(\bv{r}) = \bv{E}_0(\bv{r}) + k^2\epsilon_b \int \tensor{G}_0 (|\bv{r} - \bv{r}'|) \epsilon_C(\bv{r}') \bv{E}(\bv{r}')\, d\bv{r}'.
\label{eq:lippsch}
\end{equation}
Here, $\epsilon_C(\bv{r})$ must remain inside the integral, unlike the corresponding equation for piecewise uniform inclusions.\autocite{chen2019generalizing, bergman1980theory} The term $\bv{E}_0(\bv{r})$ is the known radiation pattern of external sources in a uniform background
\begin{equation}
\bv{E}_0(\bv{r}) = ik\int \tensor{G}_0(|\bv{r} - \bv{r}'|) Z_0\bv{J}(\bv{r}')\, d\bv{r}'.
\label{eq:E0}
\end{equation}
The Lippmann-Schwinger equation allows the field everywhere to be calculated from knowledge of the field within the interior, where $\epsilon_C(\bv{r})$ is non-zero, a key property that we exploit.

To solve \eqref{eq:lippsch}, we define an appropriate set of normal modes of the system $\bv{E}_m$, obtained by omitting source terms in \eqref{eq:inhomowave}:
\begin{equation}
\nabla\times(\nabla\times\bv{E}_m) - k^2\epsilon_b\bv{E}_m = \frac{1}{s_m}k^2\epsilon_b\epsilon_C(\bv{r})\bv{E}_m.
\label{eq:eigendiff}
\end{equation}
where $s_m$ is the $m$th eigenvalue. Alternatively, the integral form of the eigenvalue equation can be obtained from \eqref{eq:lippsch} by omitting $\bv{E}_0$,
\begin{equation}
s_m \bv{E}_m(\bv{r}) = k^2 \epsilon_b \int \tensor{G}_0 (|\bv{r} - \bv{r}'|) \epsilon_C(\bv{r}') \bv{E}_m(\bv{r}')\, d\bv{r}'.
\label{eq:eigenint}
\end{equation}
These modes are valid for the specific permittivity contrast $\epsilon_C(\bv{r})$, defined in \eqref{eq:epsC}. The eigenmodes $\bv{E}_m$ with eigenvalue $s_m$ can be regarded as an eigenmode of the permittivity profile $\epsilon_C(\bv{r})/s_m$. Note that $s_m$ can no longer be explicitly expressed as an eigenpermittivity, which was possible for a uniform interior.\autocite{chen2019generalizing}

The modes satisfy the orthogonality relation,
\begin{gather}
\int \bv{E}_n^\dagger(\bv{r}) \epsilon_C(\bv{r}) \bv{E}_m(\bv{r})\, d\bv{r} = \delta_{nm},
\label{eq:ortho}
\end{gather}
which is crucial for projection. The adjoint $\bv{E}^\dagger$ is given by the simple form
\begin{equation}
\bv{E}_m^\dagger(\bv{r}) = \bv{E}_m^\intercal(\bv{r}),
\label{eq:adjoint}
\end{equation}
rather than the Hermitian conjugate, as in the case of closed systems. The transpose here only transforms the column vector $\bv{E}(\bv{r})$ into a row vector, but does not change its spatial variation. These properties are proven in Appendix \ref{sec:ortho}.

We now proceed to solve the Lippmann-Schwinger equation \eqref{eq:lippsch} using its generalized normal modes, \eqref{eq:eigenint}. For notational brevity, we cast \eqref{eq:lippsch} in operator form,
\begin{equation}
\bv{E} = \bv{E}_0 + \hat{\Gamma}\hat{C}\bv{E},
\label{eq:lippschgamma}
\end{equation}
where $\hat{\Gamma}$ is an integral operator incorporating the Green's function along with $k^2\epsilon_b$, and $\hat{C}$ is the operator form of $\epsilon_C(\bv{r})$, so
\begin{equation}
\hat{\Gamma}\hat{C}\bv{E} \equiv k^2 \epsilon_b \int \tensor{G}_0(|\bv{r} - \bv{r}'|) \epsilon_C(\bv{r}') \bv{E}(\bv{r}')\, d\bv{r}'.
\label{eq:gammadef}
\end{equation}

The formal solution to \eqref{eq:lippschgamma} is
\begin{equation}
\bv{E} = \frac{1}{1-\hat{\Gamma}\hat{C}}\bv{E}_0.
\label{eq:formal}
\end{equation}
The solution for the unknown field $\bv{E}$ proceeds by projecting both sides, including the known $\bv{E}_0$ on to the known normal modes $\bv{E}_m$ via the projection operator
\begin{equation}
\hat{I} = \sum_m \ket{E_m}\bra{E_m}\hat{C}.
\label{eq:project}
\end{equation}
This is proved in Appendix \ref{sec:ortho} using orthogonality and modal completeness, but now cast in bra-ket notation. The unknown field $\ket{E}$ inside the inclusion is then
\begin{equation}
\ket{E} = \sum_m \ket{E_m}\bra{E_m}\frac{\hat{C}}{1-\hat{\Gamma}\hat{C}}\ket{E_0}.
\label{eq:inteqn}
\end{equation}
The operator $\hat{C}$ is scalar and commutes with other operators, allowing the operator $(1-\hat{\Gamma}\hat{C})^{-1}$ to be applied to $\bra{E_m}$, via the adjoint form of eigenvalue equation derived in \eqref{eq:eigenadj},
\begin{equation}
\bra{E_m}\hat{C}\hat{\Gamma} = \bra{E_m}s_m,
\label{eq:eigenadjop}
\end{equation}
yielding the interior field as an expansion,
\begin{equation}
\ket{E} = \sum_m \ket{E_m} \frac{1}{1-s_m} \bra{E_m}\hat{C}\ket{E_0}.
\label{eq:intsol}
\end{equation}

To obtain the fields everywhere, \eqref{eq:intsol} is inserted into the original Lippmann-Schwinger equation \eqref{eq:lippschgamma}, this time operating $\hat{\Gamma}\hat{C}$ on $\ket{E_m}$, via $s_m\ket{E_m} = \hat{\Gamma}\hat{C}\ket{E_m}$, to give
\begin{equation}
\ket{E} = \ket{E_0} + \sum_m \ket{E_m} \frac{s_m}{1-s_m} \bra{E_m}\hat{C}\ket{E_0}.
\label{eq:E0solus}
\end{equation}
For near field sources, it is more convenient to express the $\ket{E_0}$ of \eqref{eq:E0solus} in terms of $\bv{J}(\bv{r})$. This begins by casting \eqref{eq:E0} into operator form, yielding
\begin{equation}
\ket{E_0} = \frac{i}{k\epsilon_b} \hat{\Gamma} \ket{Z_0J}.
\end{equation}
After inserting into \eqref{eq:E0solus}, we obtain
\begin{equation}
\ket{E} = \ket{E_0} + \frac{i}{k\epsilon_b} \sum_m \ket{E_m} \frac{s_m}{1-s_m} \bra{E_m}\hat{C}\hat{\Gamma}\ket{Z_0J}.
\end{equation}
Again, by applying the operator $\hat{C}\hat{\Gamma}$ to $\bra{E_m}$ via \eqref{eq:eigenadjop} rather than $\ket{Z_0J}$, a simple solution is obtained 
\begin{equation}
\ket{E} = \ket{E_0} + \frac{i}{k\epsilon_b} \sum_m \ket{E_m} \frac{s_m^2}{1-s_m} \braket{E_m}{Z_0J}.
\label{eq:Jevform}
\end{equation}
Finally, the Green's tensor of the structure is obtained by choosing $\bv{J}(\bv{r})$ to be a localized Dirac-delta source and summing over all its possible orientations, which gives
\begin{equation}
\tensor{G}(\bv{r},\bv{r}') = \tensor{G}_0(|\bv{r}-\bv{r}'|) + \frac{1}{k^2\epsilon_b} \sum_m \frac{s_m^2}{(1-s_m)} \bv{E}_m(\bv{r}) \bv{E}_m^\dagger(\bv{r}'),
\label{eq:greenexp}
\end{equation}
where $\tensor{G}_0(|\bv{r}-\bv{r}'|)$ is Green's tensor of the uniform background, \eqref{eq:greensfree}, and the product $\bv{E}_m(\bv{r}) \bv{E}_m^\dagger(\bv{r}')$ yields a tensor. The solution bears remarkable similarity to the equivalent formulation for uniform inclusions.\autocite{chen2019generalizing}

\section{Obtaining the modes by re-expansion}
\label{sec:perturbation}
The bulk of the computational effort in any modal expansion method is typically devoted to finding the modes themselves. As described in the introduction, we apply techniques developed by E.\ A.\ Muljarov and others, first introduced for complex eigenfrequency modes. Similar procedures also exist for the hypbridization of modes of individual resonators to obtain modes of clusters.\autocite{bergman1980theory,rosolen2019overcoming} The target modes are expanded into the modes of a simpler open resonator. We formulate this re-expansion method to be able to obtain the modes of any finite resonator, but in this paper we shall apply it only to resonators with smoothly varying permittivity profiles where no new discontinuities are introduced. In other words, we shall use basis modes whose discontinuities coincide with the location of the discontinuities of the target geometry (see Figure \ref{fig:profile}). 

Specifically, we seek to solve the eigenvalue equation \eqref{eq:eigendiff} using modes of a system with a simpler permittivity profile $\tilde{\theta}(\bv{r})$, a Heaviside type function that is unity in the interior and zero elsewhere. In general, the non-zero region of $\tilde{\theta}(\bv{r})$ needs to enclose the non-zero region of $\epsilon_C(\bv{r})$, though these two regions may coincide as in Figure \ref{fig:profile}. The eigenvalue equation for the basis modes is identical to \eqref{eq:eigendiff}, except for $\tilde{\theta}(\bv{r})$,
\begin{equation}
\nabla\times(\nabla\times\tilde{\bv{E}}_\mu) - k^2\epsilon_b\tilde{\bv{E}}_\mu = \frac{1}{\tilde{s}_\mu}k^2\epsilon_b\tilde{\theta}(\bv{r})\tilde{\bv{E}}_\mu.
\label{eq:eigenb}
\end{equation}
Tildes are affixed to quantities specifically associated with the basis modes, which are also labeled by Greek indexes. Such modes were discussed in more detail in Ref.\ \parencite{chen2019generalizing}. The modes of \eqref{eq:eigenb} also obey the important orthonormality relation
\begin{gather}
\int \tilde{\bv{E}}_\nu^\dagger(\bv{r}) \tilde{\theta}(\bv{r}) \tilde{\bv{E}}_\mu(\bv{r})\, d\bv{r} = \delta_{\nu\mu},
\label{eq:orthob}
\end{gather}
derived in Appendix \ref{sec:ortho}. Likewise, its adjoint mode $\tilde{\bv{E}}^\dagger_\nu$ is given by \eqref{eq:adjoint}.

We mention one important disparity between the target modes of \eqref{eq:eigendiff} and basis modes of \eqref{eq:eigenb}. In the latter, the interior permittivity profile is uniform, and the divergence here is zero, $\nabla\cdot\tilde{\bv{E}}_\mu = 0$. But in the former, a changing permittivity profile within the interior implies non-zero divergence, $\nabla\cdot\bv{E}_m \neq 0$. It would thus seem that the basis modes defined by \eqref{eq:eigenb} are insufficient to represent the target modes of \eqref{eq:eigendiff}, but a closer analysis, provided in Section \ref{sec:basis}, reveals how this is possible.

\subsection{Derivation of matrix eigenvalue equation}
\label{sec:deriv}
Our derivation follows the familiar pattern of expanding the target differential equation using a complete basis,\autocite{morse1946methods} initially with unknown coefficients \eqref{eq:expansion}. We then use orthonormal projection to evaluate the coefficients, resulting in a set of integrals over the perturbation that populate a linear algebraic system of equations, which is solved for the final solution. The formulation developed can be used to find the modes of any permittivity profile $\epsilon_C(\bv{r})$ enclosed by the interior of $\tilde{\theta}(\bv{r})$ from \eqref{eq:eigenb}. This includes the case where the outer surface of $\epsilon_C(\bv{r})$ differs from $\tilde{\theta}(\bv{r})$, but we shall not treat this case in this paper.

Inserting \eqref{eq:expansion} into \eqref{eq:eigendiff} gives
\begin{equation}
\sum_\mu c_{\mu,m} [\nabla\times(\nabla\times\tilde{\bv{E}}_\mu) - k^2\epsilon_b\tilde{\bv{E}}_\mu] = \sum_\mu \frac{c_{\mu,m}}{s_m}k^2\epsilon_b\epsilon_C(\bv{r})\tilde{\bv{E}}_\mu.
\end{equation}
We notice that the LHS satisfies the eigenvalue equation of the basis modes \eqref{eq:eigenb}, therefore
\begin{equation}
\sum_\mu \frac{c_{\mu,m}}{\tilde{s}_\mu} \tilde{\theta}(\bv{r}) \tilde{\bv{E}}_\mu = \sum_\mu \frac{c_{\mu,m}}{s_m} \epsilon_C(\bv{r}) \tilde{\bv{E}}_\mu,
\label{eq:eigenexpanded}
\end{equation}
The unknown coefficients are found by projecting onto the basis modes by taking the scalar product, over the regions where $\tilde{\theta}(\bv{r})$ and $\epsilon_C(\bv{r})$ are respectively non-zero,
\begin{equation}
\sum_\mu \frac{c_{\mu,m}}{\tilde{s}_\mu} \int \tilde{\bv{E}}^\dagger_\nu \tilde{\theta}(\bv{r}) \tilde{\bv{E}}_\mu\, d\bv{r} = \sum_\mu \frac{c_{\mu,m}}{s_m} \int \tilde{\bv{E}}^\dagger_\nu \epsilon_C(\bv{r}) \tilde{\bv{E}}_\mu\, d\bv{r}.
\label{eq:pertderiv}
\end{equation}
The integral on the LHS can be simplified using the orthogonality relation \eqref{eq:orthob} to yield $\delta_{\nu\mu}$. Then, by defining the matrix element for the RHS
\begin{equation}
V_{\nu\mu} = \int \tilde{\bv{E}}^\dagger_\nu(\bv{r}) \epsilon_C(\bv{r}) \tilde{\bv{E}}_\mu(\bv{r})\, d\bv{r},
\label{eq:Vdef}
\end{equation}
we obtain a linear matrix eigenvalue problem
\begin{equation}
s_m c_{\nu,m} = \tilde{s}_\nu \sum_\mu V_{\nu\mu} c_{\mu,m}.
\label{eq:perteig}
\end{equation}

Solution of \eqref{eq:perteig} yields the target modes $\bv{E}_m$ and their eigenvalues $s_m$ for the geometry defined by $\epsilon_C(\bv{r})$, expanded in terms of the basis modes $\tilde{\bv{E}}_\mu$. The numerical implementation of \eqref{eq:perteig} first requires preparation of all the basis modes, including a set of longitudinal modes, the latter of which is discussed in detail in Section \ref{sec:basis}. Next, the integrals of each matrix element of \eqref{eq:Vdef} must be computed, and then \eqref{eq:perteig} is ready to be solved, using any numerical linear algebra package for small dense systems. For numerical efficiency, the system of equations \eqref{eq:perteig} can be symmeterized, a process considered below. After the solution of \eqref{eq:perteig}, normalization completes the process, a process described in Section \ref{sec:norm}. Finally, the solution of the scattering problem is computed via GENOME, \eqref{eq:greenexp} or \eqref{eq:E0solus}. The latter requires overlap integrals $\bra{E_m}\hat{C}\ket{E_0}$ to be evaluated.

Symmeterization of \eqref{eq:perteig} may enable the use of a more efficient linear algebra routine. The only impediment is the factor $\tilde{s}_\nu$ in \eqref{eq:perteig}, since the matrix $V_{\nu\mu}$ is symmetric by virtue of the simple adjoint \eqref{eq:adjoint}. To remedy this, we need only divide and multiply appropriately by $\sqrt{\tilde{s}}$,
\begin{equation}
s_m \left[\sqrt{\frac{s_m}{\tilde{s}_\nu}} c_{\nu,m}\right] = \sum_\mu \left[\sqrt{\tilde{s}_\nu} V_{\nu\mu} \sqrt{\smash{\tilde{s}_\mu}\vphantom{\tilde{s}_\nu}}\right] \left[\sqrt{\frac{s_m}{\tilde{s}_\mu}} c_{\mu,m}\right].
\label{eq:symeig}
\end{equation}
We have also multiplied by $\sqrt{s_m}$, as this enables automatic normalization. This matrix eigenvalue problem can be solved for $b_{\mu,m} \equiv \sqrt{s_m/\tilde{s}_\mu} c_{\mu,m}$, from which the standard coefficients can be retrieved.

\subsection{Normalization and orthogonality}
\label{sec:norm}
All modal expansion methods require the modes be normalized, as this enables projection. We show that a convenient evaluation of \eqref{eq:ortho} is possible if the basis modes $\{\tilde{\bv{E}}_\mu\}$ are already normalized according to \eqref{eq:orthob}. We expand the normalization integral \eqref{eq:ortho} using the basis modes \eqref{eq:expansion},
\begin{equation}
\begin{aligned}
\int \bv{E}_n^\dagger(\bv{r}) \epsilon_C(\bv{r}) \bv{E}_m(\bv{r})\, d\bv{r} &= 
\sum_\nu\sum_\mu c_{\nu,n} V_{\nu\mu} c_{\mu,m}\\
&= s_m \sum_\mu \frac{c_{\mu,n} c_{\mu,m}}{\tilde{s}_\mu}\\
&= \sum_\mu b_{\mu,n} b_{\mu,m} = \delta_ {nm}
\end{aligned}
\label{eq:normderiv}
\end{equation}
To obtain the result, we required the definition of $V_{\nu\mu}$ in \eqref{eq:Vdef} and the eigenvalue equation \eqref{eq:perteig}. Finally, we have inserted the symmeterized coefficients $b_{\mu,m}$ defined in \eqref{eq:symeig} and assumed that $\sum_\mu b_{\mu,m}^2=1$, since this is precisely the dot product between the normalized left and right eigenvectors of the symmeterized matrix operator in \eqref{eq:symeig}, which is enforced by some, though not all, linear algebra packages.

One interesting consequence of \eqref{eq:normderiv} is that orthogonality is numerically satisfied to machine precision regardless of truncation, since the eigenvectors of a complex symmetric matrix \eqref{eq:symeig} obey the complex orthogonality relation, $\sum_\mu b_{\mu,n} b_{\mu,m} = \delta_ {nm}$. The exception is modes belonging to the same eigenvalue, such as symmetry degenerate modes, which sometimes need to be orthogonalized manually depending on the linear algebra package.

\section{Longitudinal basis modes}
\label{sec:basis}
The success of the re-expansion method relies on the completeness of the basis modes \eqref{eq:eigenb}. It is therefore useful to understand the types of modes that \eqref{eq:eigenb} admits,\autocite{bergman1980theory} of which we shall use two: transverse and longitudinal electric. For physical insight, it is useful first to reinterpret the eigenvalue in \eqref{eq:eigenb} as an eigenpermittivity, $\tilde{s}_\mu = \epsilon_b/(\tilde{\epsilon}_\mu-\epsilon_b)$.\autocite{chen2019generalizing} Then, the eigenvalue equation has the simple form in its uniform interior,
\begin{equation}
\nabla\times(\nabla\times\tilde{\bv{E}}_\mu) - k^2\tilde{\epsilon}_\mu\tilde{\bv{E}}_\mu = 0.
\label{eq:eigensplit}
\end{equation}
Here, the modes are divergence-free, which can be seen by applying the divergence operator to each side of \eqref{eq:eigensplit}, obtaining $\nabla\cdot\tilde{\bv{E}}_\mu = 0$. They are not divergence-free along the boundary between the interior and the exterior, but we shall nevertheless call them transverse modes. Alone, these modes are not sufficient to represent the target modes. For completeness, we briefly overview these transverse basis modes in Appendix \ref{sec:transverse}.

More sets of modes arise when the eigenvalue is $\tilde{s}_\mu = -1$ or $\tilde{\epsilon}_\mu = 0$. The eigenvalue equation \eqref{eq:eigenb} then reduces to
\begin{align}
\nabla\times(\nabla\times\tilde{\bv{E}}_\mu) = 0,
\label{eq:longeigen}
\end{align}
in the interior. Since setting $\tilde{\epsilon}_\mu = 0$ is mathematically indistinguishable from setting $k=0$, these modes behave as static fields. These modal fields are identically zero in the exterior since the index contrast with the background is infinite. In Section \ref{sec:Lmodes}, we show that longitudinal electric modes are generated from the subset of \eqref{eq:longeigen} where $\nabla\times\tilde{\bv{E}}_\mu = 0$, providing the necessary longitudinal component for representing the target modes. Another set of modes is generated by \eqref{eq:longeigen}, featuring magnetostatic fields. But since we treat non-magnetic media, these modes are not required, and we shall hereafter refer to longitudinal electric modes as simply longitudinal modes.

\subsection{Definitions and properties}
\label{sec:Lmodes}
Longitudinal modes are seldom seen or used in the literature, so we devote this section to discussing their properties. We intend for the section to be introductory, so we expand slightly upon other expositions.\autocite{bergman1980theory,lobanov2019resonant} We treat the longitudinal electric set of modes that arises from \eqref{eq:longeigen}, whereby $\nabla\times\tilde{\bv{E}}_\mu = 0$.

Firstly, irrotational vector fields can be expressed in terms of potential
\begin{equation}
\tilde{\bv{E}}_\mu = \nabla\tilde{\phi}_\mu.
\label{eq:potendef}
\end{equation}
Since $\tilde{\epsilon}_\mu = 0$, there is infinite index contrast between the interior and exterior, and the field is identically zero outside. In order to satisfy the boundary conditions at the interface, the parallel component of the interior electric field must be zero here, but the perpendicular component can be arbitrary. The first condition implies that
\begin{equation}
\nabla_\parallel\tilde{\phi}_\mu = \tilde{\bv{E}}_{\mu,\parallel} = 0,
\end{equation}
along the boundary, so $\tilde{\phi}_\mu$ is constant there. This is directly analogous to the external surface of a perfect conductor being an equipotential surface. Due to the freedom associated with the definition \eqref{eq:potendef}, we may add any constant to $\tilde{\phi}_\mu$ without affecting $\tilde{\bv{E}}_\mu$. This allows us to adjust $\tilde{\phi}_\mu$ such that they satisfy Dirichlet boundary conditions
\begin{equation}
\tilde{\phi}_\mu\biggr\rvert_{\partial B} = 0,
\label{eq:dirichlet}
\end{equation}
where $\partial B$ defines the interface between the interior and exterior.

The boundary condition \eqref{eq:dirichlet} is sufficient to demonstrate that these modes are longitudinal. If $\nabla^2\tilde{\phi}_\mu$ is everywhere zero, then $\tilde{\phi}_\mu$ is also everywhere zero, and $\tilde{\bv{E}}_\mu$ would be a constant, a trivial solution. Otherwise, if $\nabla^2\tilde{\phi}_\mu = \nabla\cdot\tilde{\bv{E}}_\mu$ is not everywhere zero, $\tilde{\bv{E}}_\mu$ must be longitudinal. So if non-trivial solutions exist, they must be longitudinal.

Until now, we have shown that longitudinal modes are subject only to the constraints \eqref{eq:potendef} and \eqref{eq:dirichlet}. So great freedom exists in their construction, and their functional forms can be almost arbitrary. To be useful, we must construct a complete and orthonormal basis of longitudinal modes for use in the expansion \eqref{eq:expansion}. This can be achieved by artificially imposing the additional constraint that the potentials are eigenmodes of the Laplace operator, 
\begin{equation}
\nabla^2 \tilde{\phi}_\mu + \alpha_\mu^2 \tilde{\phi}_\mu = 0
\label{eq:Lhelmholtz}
\end{equation}
This corresponds to a set of cavity modes of a closed electrostatic resonator satisfying the Dirichlet boundary condition \eqref{eq:dirichlet}. 
Such modes can be shown to be everywhere non-transverse by explicit computation,
\begin{equation}
\nabla\cdot\tilde{\bv{E}}_\mu = \nabla^2\tilde{\phi}_\mu = -\alpha_\mu^2\tilde{\phi}_\mu \neq 0.
\end{equation}
The form of \eqref{eq:Lhelmholtz} can be justified by an alternative derivation: we generate a complete orthonormal basis for separable geometries by demanding that $\tilde{\phi}_\mu$ be orthogonal along each separable coordinate. This is the procedure considered in Section \ref{sec:Lfourier}, and leads to a basis based on generalized Fourier series. The Fourier-Bessel series is most efficient for smooth functions, and is suitable for smooth perturbations, and can even allow the re-expansion to treat anisotropic inclusions.

Finally, longitudinal basis modes must be normalized according to \eqref{eq:orthob},
\begin{equation}
\begin{aligned}
\int \tilde{\bv{E}}_\nu^\dagger \tilde{\bv{E}}_\mu\, d\bv{r} &= \int \nabla\tilde{\phi}^\dagger_\nu\cdot\nabla\tilde{\phi}_\mu\, d\bv{r}\\
&= \oint\tilde{\phi}^\dagger_\nu\nabla\tilde{\phi}_\mu\cdot d\bv{S} - \int\tilde{\phi}^\dagger_\nu\nabla^2\tilde{\phi}_\mu\, d\bv{r}\\ 
&= \alpha_\mu^2\int\tilde{\phi}_\nu^\dagger\tilde{\phi}_\mu\, d\bv{r}.
\end{aligned}
\label{eq:Lnormold}
\end{equation}
In all integrals, the domain of integration is the resonator interior, in accordance with \eqref{eq:orthob}. The surface integral vanishes due to the boundary condition \eqref{eq:dirichlet}, while \eqref{eq:Lhelmholtz} is used to simplify the volume integral. Thus, normalization can be reduced to integrals over potentials. 

\subsection{Fourier-Bessel basis of longitudinal modes}
\label{sec:Lfourier}
For circular and spherical geometries, an orthonormal basis of longitudinal modes can be constructed, yielding the Fourier-Bessel series. We may then deduce the generalization to arbitrary shapes. For example, we provide a prescription for a 2D circular domain of radius $B$, deriving the basis by demanding orthogonality in each direction. We employ polar coordinates, $(r,\vartheta)$. In the angular direction, the continuous rotational symmetry furnishes the well known orthogonal set of cylindrical harmonics, $e^{i\tau\vartheta}$, or equivalently $\{\sin(\tau\vartheta), \cos(\tau\vartheta)\}$. In the radial direction, orthogonality is expressed as
\begin{equation}
\int_0^B \tilde{\phi}^\dagger_\nu \tilde{\phi}_\mu r\, dr = \delta_{\nu\mu}
\label{eq:orthofn}
\end{equation}
subject to the known boundary conditions that $\{\tilde{\phi}_\mu\}$ all be zero at $r=B$ and be bounded at $r=0$. Although \eqref{eq:orthofn} enforces orthogonality only among the potentials, it leads to orthogonality among the modes \eqref{eq:Lnormold} once we justify \eqref{eq:Lhelmholtz}. From \eqref{eq:orthofn}, a set of orthogonal functions or a generalized Fourier series is defined, where $r$ is the weight function, corresponding to the Bessel functions.\autocite{lobanov2019resonant,kreyszig2011advanced} The overall solution may now be constructed,
\begin{equation}
\tilde{\phi}_{\tau\tau'}^{\{o,e\}} = L_{\tau\tau'} J_\tau\left(\frac{u_{\tau\tau'}}{B} r\right)
\genfrac{\lbrace}{\rbrace}{0pt}{}{\sin(\tau\vartheta)}{\cos(\tau\vartheta)},
\label{eq:FBmode}
\end{equation}
where we have employed two integer subscripts $\mu \equiv (\tau,\tau')$, the normalization constant is $L_{\tau\tau'}$, and $u_{\tau\tau'}$ is the $\tau'$-th root of Bessel function $J_\tau(z)$. Superscripts are used to denote the odd or even solution. This solution is valid for the interior of the resonator only, as the potential is identically zero outside. We may now observe that the construction \eqref{eq:FBmode} obeys the Helmholtz equation \eqref{eq:Lhelmholtz} with Dirichlet boundary conditions \eqref{eq:dirichlet}, identifying $\alpha_\mu = u_{\tau\tau'}/B$. The importance of \eqref{eq:Lhelmholtz} is that it generalizes \eqref{eq:FBmode} to arbitrary geometries of any dimension, where the procedure of enforcing orthogonality along each dimension is impossible.

Since \eqref{eq:FBmode} satisfies the Helmholtz equation \eqref{eq:Lhelmholtz}, \eqref{eq:Lnormold} can be used to normalize the modes. The radial part can be evaluated analytically via the defining orthogonality relation of Bessel functions,
\begin{equation}
\int_0^1 J_\tau(u_{\tau\tau'}r) J_\tau(u_{\tau\sigma'}r)r\, dr = \frac{\delta_{\tau'\sigma'}}{2} J_{\tau+1}^2(u_{\tau\tau'}),
\end{equation}
to yield
\begin{equation}
L_{\tau\tau'} = [\sqrt{\pi} \alpha_{\tau\tau'} B J_{\tau+1}(\alpha_{\tau\tau'} B)]^{-1}.
\end{equation}
Together with \eqref{eq:FBmode}, this defines the desired orthonormal set of longitudinal modes for circular domains.

\section{Numerical results}
\label{sec:numerics}
\subsection{Implementation}
Before presenting the results of our numerical implementation, we first recapitulate the process of re-expansion to obtain the target eigenmodes of the resonator with a smoothly varying permittivity profile specified by a target equation \eqref{eq:eigendiff}, and then using these modes to solve Maxwell's equations using GENOME. 

Finding the target modes begins by obtaining all necessary basis modes (Figure \ref{fig:profile}). The transverse basis modes and their eigenvalues are defined for the equivalent uniform resonator, found by solving the eigenvalue equation \eqref{eq:eigenb}. The longitudinal basis modes are found from \eqref{eq:Lhelmholtz}. If the basis structure is simple like the examples considered in this paper, then analytical results are available for these two steps, such as \eqref{eq:FBmode} and Appendix \ref{sec:transverse}. Otherwise, numerical methods may be used to obtain the basis modes. Since only a finite number of basis modes can be used, a truncation procedure is required, beginning with modes that contain the least number of nodes, which usually corresponds to modes with the smallest absolute eigenpermittivity, $|\tilde{\epsilon}_\mu| = |1/\tilde{s}_\mu+1|$. The optimal ratio of transverse and longitudinal basis is often not known, and should be studied in convergence tests, to be further detailed in Section \ref{sec:modes}. Then, the overlap integrals among the chosen basis modes \eqref{eq:Vdef} are evaluated, where the adjoint basis modes are defined by \eqref{eq:adjoint}. This forms the matrix eigenvalue equation \eqref{eq:perteig}, which optionally may be symmeterized to give \eqref{eq:symeig}. Its numerical solution yields the eigenvalues of the target modes and their fields, expanded using the basis modes, according to \eqref{eq:expansion}. This completes the process of obtaining the target modes, the solutions of \eqref{eq:eigendiff}.

The eigenmodes are automatically normalized when solving the symmetrized form \eqref{eq:symeig}. To use GENOME, \eqref{eq:Jevform} is applied to calculate the total fields produced by any given source configuration. This involves the evaluation of overlap integrals between the source and the adjoint modes and knowledge of the incident field. Alternatively, the Green's tensor of the resonator can be obtained using \eqref{eq:greenexp}, which requires the known Green's tensor of free space.

\begin{figure}[!t]
\begin{center}
\includegraphics{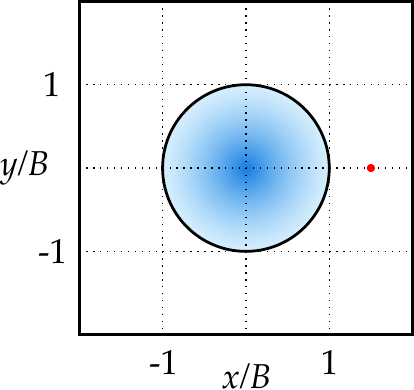}
\caption{Schematic of the graded index structure used for numerical examples. A 2D cylinder of radius $B$ with a parabolic permittivity profile is placed in an infinite background. It is excited by a point dipole whose location is indicated by the red dot, in the examples of Section \ref{sec:fields}. The rectangular border is the limit of the plotting domain for the examples in Section \ref{sec:fields}, while a slightly smaller domain is used in Section \ref{sec:modes}.}
\label{fig:geometry}
\end{center}
\end{figure}

\subsection{Simulation geometry}
In the sections to follow, we present the numerical results from our implementation. We choose to consider the 2D example shown in Figure \ref{fig:geometry}, consisting of a graded index fiber of radius $B$ embedded in an infinite background of permittivity $\epsilon_b = 1$. The system is excited by a point dipole oscillating at a fixed frequency $kB=1$. Lengths are specified in arbitrary units, since the units of wavenumber $k$ are given in units of inverse length.

In this case, the basis modes can be found by solving the step-index fiber dispersion relation \eqref{eq:disprel} of Appendix \ref{sec:transverse}, a transcendental equation that can be quickly and reliably solved using the argument principle method using freely available code.\autocite{chen2017robust} As shown in Figure \ref{fig:profile}, our target structure has a parabolic axisymmetric index profile, with a permittivity contrast defined by $\epsilon_C(\bv{r}) = 2-(r/B)^2$ within the fiber interior $(r < B)$, so that the structure has relative permittivity 3 at its center, 2 at its interface, and 1 within the background. We chose an axisymmetric profile since they are the most widely applicable, but we emphasize that any smooth index profile can be treated by the present method, whether axisymmetric or not. This includes the case where the discontinuity in $\epsilon_C(\bv{r})$ varies along the exterior surface. In Section \ref{sec:modes}, we present the modes of this structure, along with the convergence properties of the re-expansion method, while in Section \ref{sec:fields}, we obtain Green's tensor of the structure using GENOME \eqref{eq:greenexp}. 

\subsection{Modes of graded index structure}
\label{sec:modes}
\begin{figure}[!t]
\begin{center}
\subfloat[$s_m = 0.287563463191829 + 0.107337071161170i$]{\begin{tabular}[b]{c}%
	\includegraphics{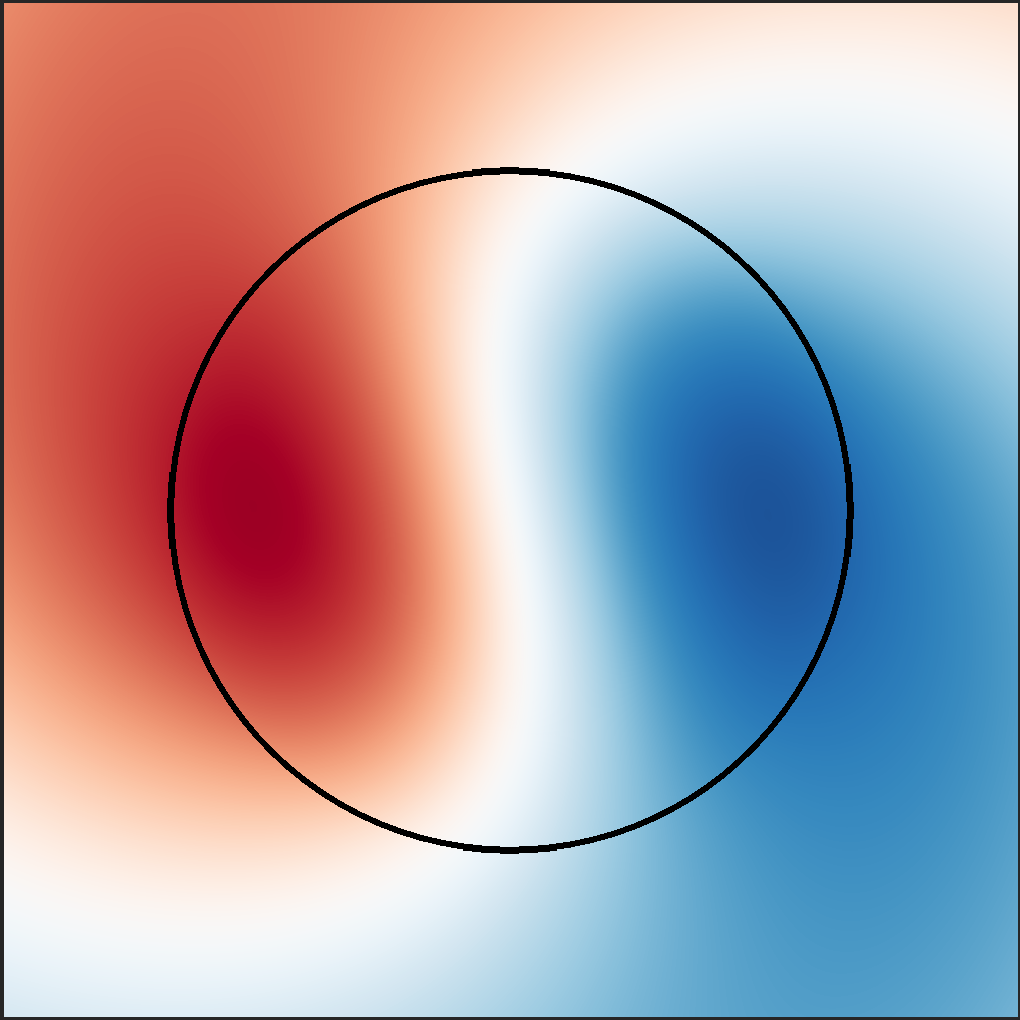}
	\end{tabular}}
\subfloat[$s_m = 0.055285453048475 + 0.003657335781741i$]{\begin{tabular}[b]{c}%
	\includegraphics{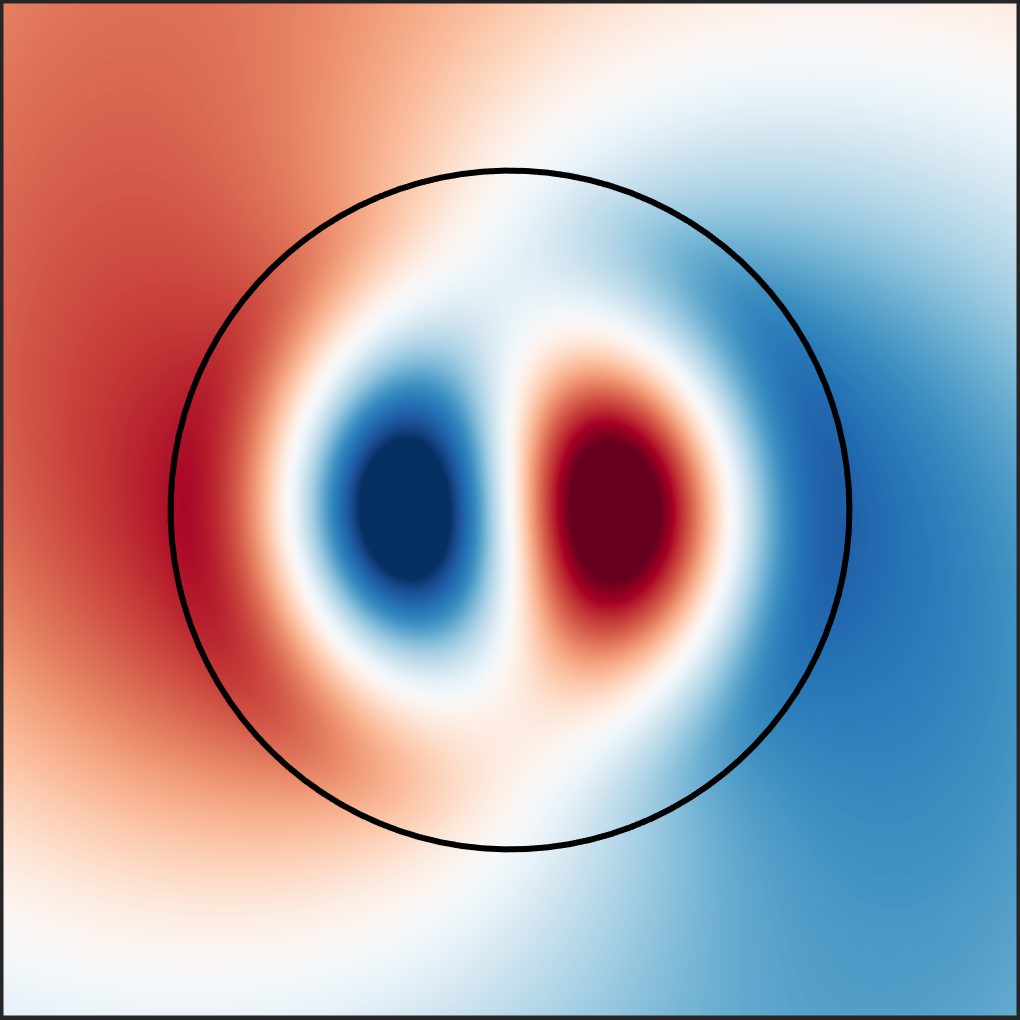}
	\end{tabular}}
\subfloat{
	\includegraphics{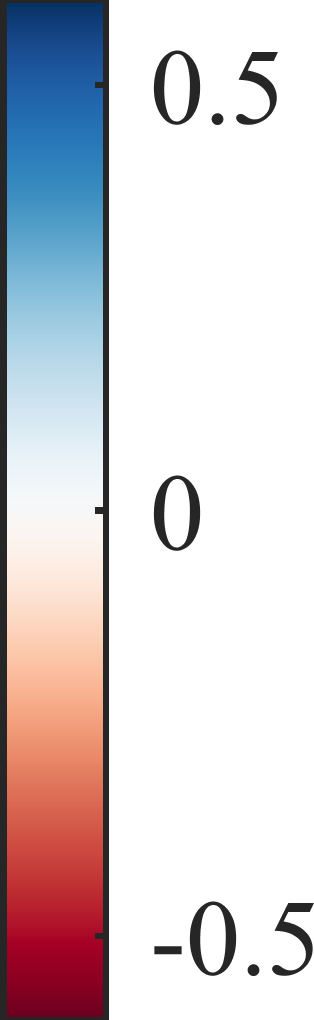}}
\caption{The two fundamental TM modes of the graded index structure of azimuthal order $\tau=1$, plotting the $\real(E_z)$ component.}
\label{fig:TMmodes}
\end{center}
\end{figure}

\begin{figure}[!t]
\begin{center}
\subfloat[$s_m = -0.659312291068941 + 0.431135132638932i$]{\begin{tabular}[b]{c}%
	\includegraphics{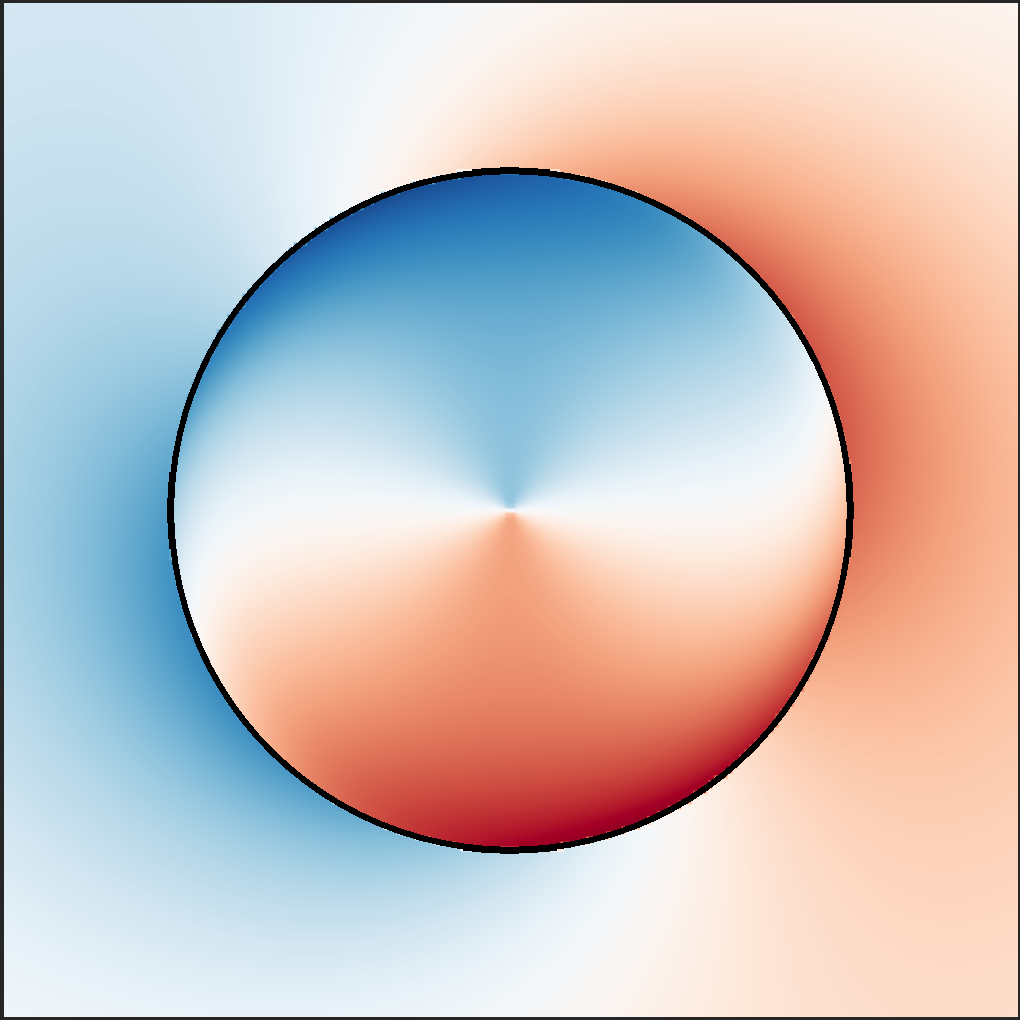}
	\end{tabular}}
\subfloat[$s_m = 0.119461090265710 + 0.016012447606085i$]{\begin{tabular}[b]{c}%
	\includegraphics{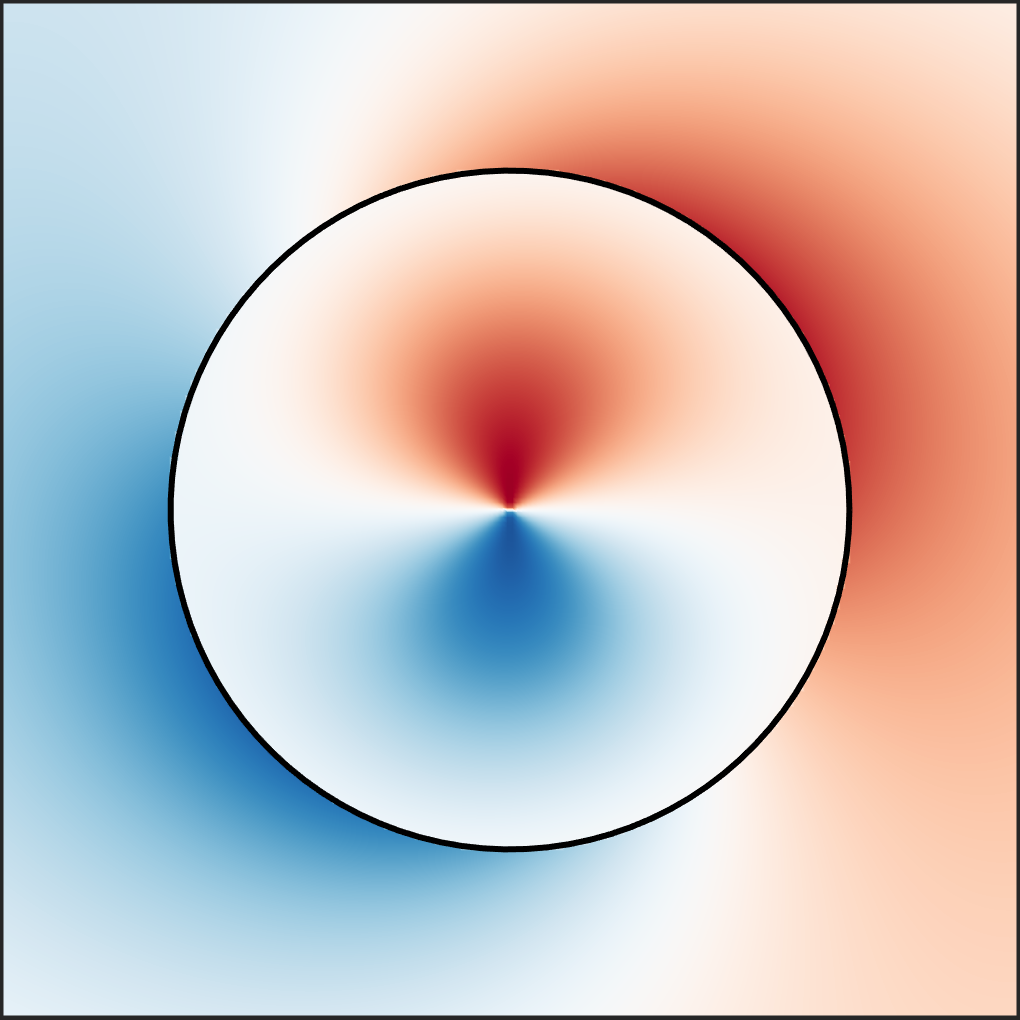}
	\end{tabular}}
\subfloat{
	\includegraphics{modebar.png}}
\caption{The two fundamental TE modes of the graded index structure of azimuthal order $\tau=1$, plotting the $\real(E_r)$ component.}
\label{fig:TEmodes}
\end{center}
\end{figure}

Our chosen geometry possesses several useful symmetries. Firstly, cylindrical symmetry means that all basis and target modes can be assigned an azimuthal and a radial quantum number, representing the number of nodes each mode has along the respective directions. For example, for the longitudinal basis modes, this corresponds precisely to the two indexes $\tau$ and $\tau'$ of the Fourier-Bessel series \eqref{eq:FBmode}. In this case, basis modes of different azimuthal orders never interact with each other, and the corresponding matrix elements \eqref{eq:Vdef} are identically zero. However, basis modes of the same azimuthal order but different radial order do interact to yield the target modes. This allows the matrix eigenvalue equation \eqref{eq:perteig} to be put into block diagonal form, solving each block separately.

Since the structure is 2D, all modes can be categorized as either transverse magnetic (TM) or transverse electric (TE). TM and TE basis modes never interact since their matrix elements \eqref{eq:Vdef} are zero for any structure with an isotropic permittivity tensor. Thus, the TM modes of the graded index structure are represented by only TM basis modes, and likewise for TE modes. The TM case is simpler since the electric field is exclusively out of plane, which we denote by $E_z$. Since this is the only electric field component, fields are still strictly transverse with $\nabla\cdot\bv{E} = 0$, despite the spatial variation of $\epsilon_C(\bv{r})$. This means that no longitudinal basis modes are necessary for expansion, as all longitudinal modes in 2D have zero $E_z$.

\begin{figure}[!t]
\begin{center}
\includegraphics[width=8.5cm]{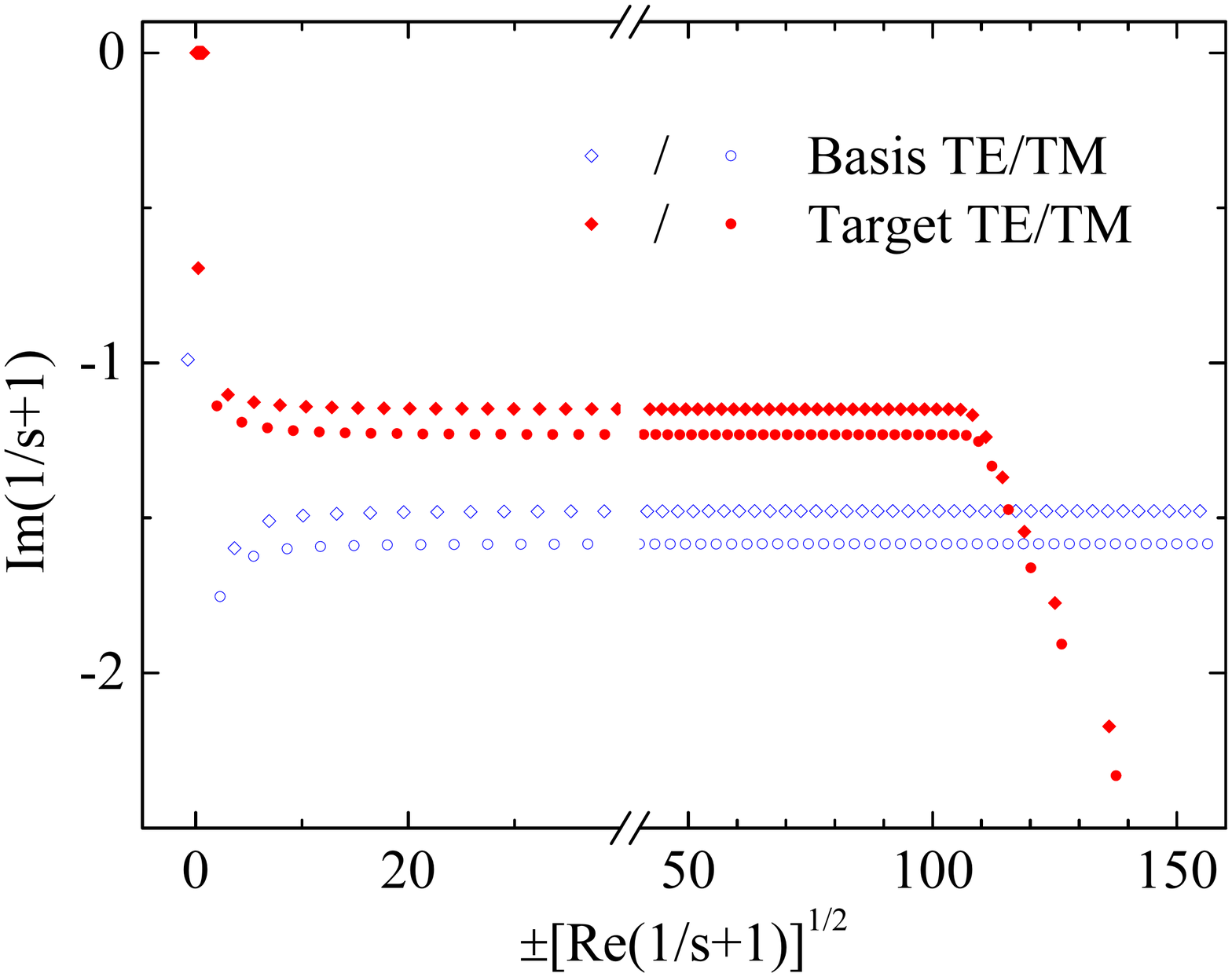}
\caption{Eigenvalues of the modes of the graded index structure (solid red) and their constituent basis modes (hollow blue). TM modes are indicated by round markers, and TE modes by diamonds. Plotted are the real and imaginary parts of the inverse of the eigenvalues plus unity, $1/s+1$. The horizontal axis has been compressed by plotting the square root, $\sqrt{\real(1/s+1)}$. The sole exception is the leftmost TE basis mode, which is negative, for which we plot $-\sqrt{-\real(1/s+1)}$.}
\label{fig:eigenvalues}
\end{center}
\end{figure}

Considering first the simpler TM case, Figure \ref{fig:TMmodes} shows the two fundamental modes, for azimuthal order $\tau=1$, simulated using the 300 lowest TM transverse basis modes ordered by absolute eigenpermittivity. The computed eigenvalues are displayed alongside each mode. The $\real(E_z)$ component is plotted, as this is the only electric field component. For convenience, we have constructed the form of the mode with angular dependence $e^{i\tau\vartheta}$ from the degenerate pair with angular dependence $\{\cos(\tau\vartheta), \sin(\tau\vartheta)\}$. This form is easier to visualize, as the real and imaginary parts of the modal fields are $90^\circ$ rotations of each other, so only one plot is required to display $E_z$. Higher order radial modes of each azimuthal order have also been generated by the method, though their accuracies progressively decrease, especially for modes near the truncation limit (see Figure \ref{fig:eigenvalues}). This is because higher order target modes tend to require higher order basis modes, which may be beyond the truncation limit. In Figures \ref{fig:TMmodes} (a) and (b), the modal fields differ only within the interior. Indeed, all modes of the same azimuthal order are identical within the background up to a multiplicative factor. 

Consider now the TE modes of the graded index structure, which do require the longitudinal basis modes. Indeed, failure to include longitudinal modes means that the target modes will never be produced correctly regardless of the number of transverse basis modes used. Displayed in Figure \ref{fig:TEmodes} are the two fundamental modes for $\tau=1$. The modes were obtained using 300 transverse and 300 longitudinal basis modes. The $\real(E_r)$ component is shown, as this component exhibits the discontinuity at the boundary of the cylinder. Again, the $e^{i\tau\vartheta}$ form of the modes is displayed. 

We now compare the eigenvalues of the target modes of azimuthal order $\tau = 1$, both TM and TE, against the eigenvalues of their constituent transverse basis modes in Figure \ref{fig:eigenvalues}. The eigenvalues of the longitudinal basis modes are not displayed, since they are all $\tilde{s}_\mu = -1$. We choose to plot the inverse of eigenvalues, more specifically $1/s + 1$, for two reasons. Firstly, this corresponds to the eigenpermittivity, $\tilde{\epsilon}_\mu = 1/\tilde{s}_\mu + 1$, in the case of the basis modes \eqref{eq:eigenb}. Secondly, a more recognizable trend is observed, whereby successively higher order modes have larger real parts but similar imaginary parts. The target modes to the right of the figure do not follow this trend, since these are increasingly erroneous due to truncation.

\begin{figure}[!t]
\begin{center}
\includegraphics{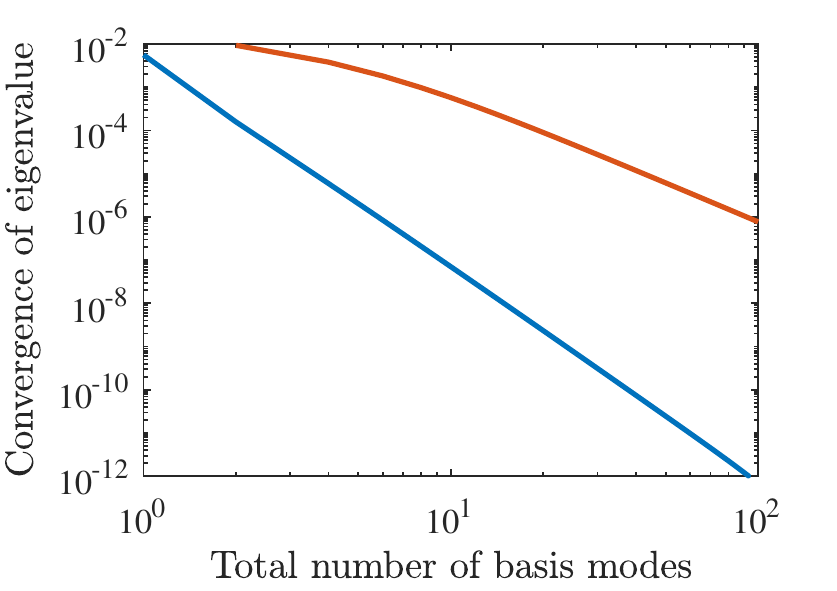}
\caption{Convergence in eigenvalue of the fundamental TM mode (blue) of Figure \ref{fig:TMmodes} (a), and the fundamental TE mode (red) of Figure \ref{fig:TEmodes} (a). Displayed is the relative difference in the computed eigenvalue with respect to a reference value computed with 300 basis modes (blue) and 600 basis modes (red), as a function of number of basis modes. No longitudinal modes are used for the TM mode (blue), as they are unnecessary, while data for the TE mode (red) is for a fixed $1:1$ ratio of transverse to longitudinal modes, as indicated in Figure \ref{fig:TEconv}.}
\label{fig:conv}
\end{center}
\end{figure}

\begin{figure}[!t]
\begin{center}
\includegraphics{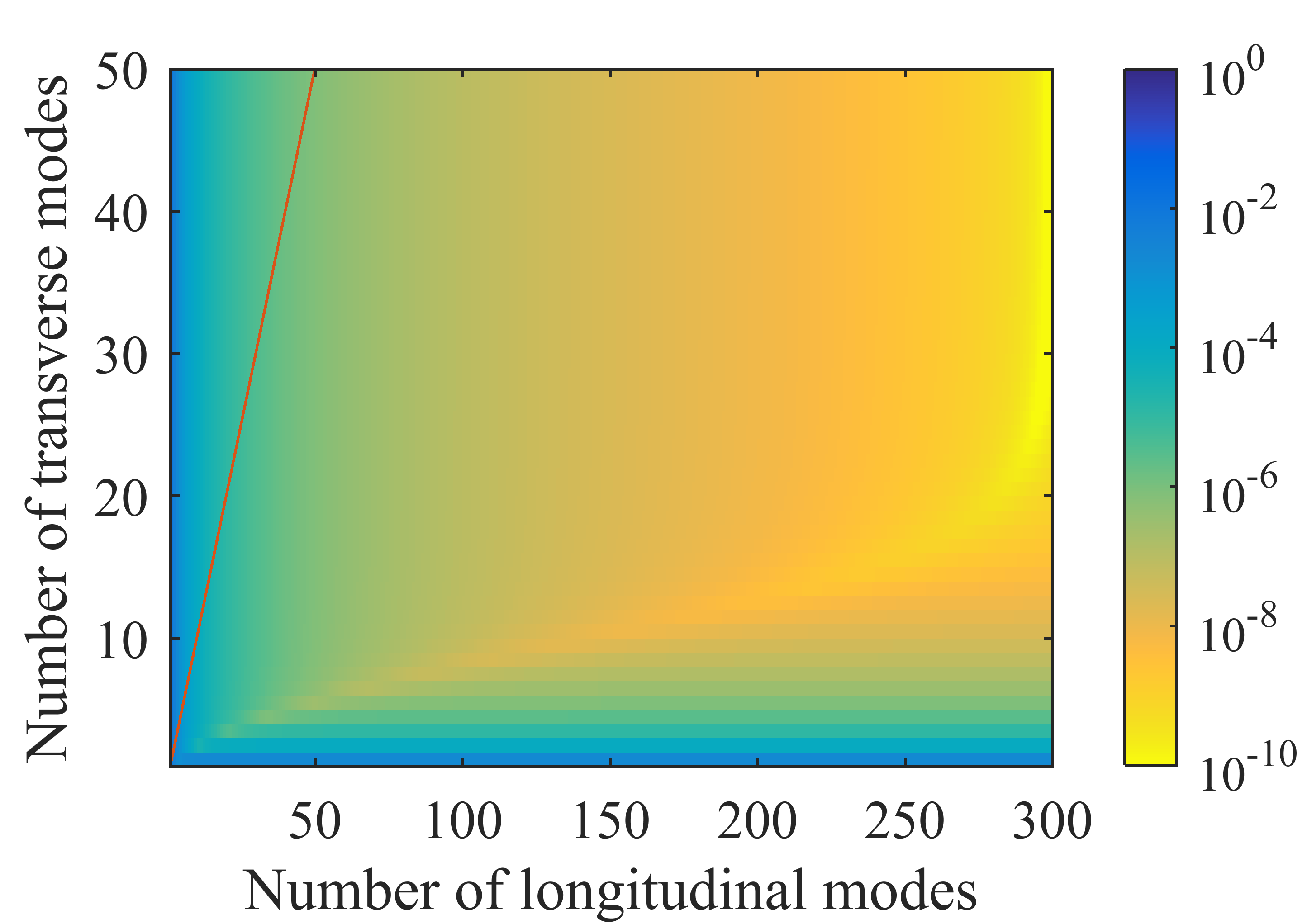}
\caption{Convergence in eigenvalue of the fundamental TE mode shown in Figure \ref{fig:TEmodes} (a). Shown is the relative difference in the computed eigenvalue with respect to a reference value computed with 300 transverse and 300 longitudinal basis modes, as a function of number of transverse and longitudinal basis modes. The color scale displays the relative difference with respect to the reference value. The red line indicates the trajectory used for the red line in Figure \ref{fig:conv}.}
\label{fig:TEconv}
\end{center}
\end{figure}

The re-expansion method shows rapid convergence properties. The behavior of the TM and TE modes differ, largely because longitudinal modes are needed for the TE modes. Considering first the simpler TM case, we treat the convergence of the fundamental mode shown in Figure \ref{fig:TMmodes} (a). Displayed in Figure \ref{fig:conv} is the relative difference in the computed eigenvalue with respect to a highly accurate reference value. With less than 10 modes, excellent convergence is obtained. This is likely because the fundamental mode of the graded index structure resembles the fundamental mode of the equivalent uniform structure. Plotting the data of Figure \ref{fig:conv} on a log-log scale reveals that convergence of TM modes goes as $N^{-5}$ with the number of modes.

The convergence of the fundamental TE mode of Figure \ref{fig:TEmodes} is shown in Figure \ref{fig:TEconv} with number of transverse and longitudinal modes. We duplicate the data of Figure \ref{fig:TEconv} in Figure \ref{fig:conv}, for a fixed $1:1$ ratio of transverse to longitudinal modes. We selected this ratio to display because other ratios do not produce significantly better convergence within this data range. Convergence is still rapid, but in comparison to the TE case, many longitudinal modes are now required to achieve similar accuracy. This is likely because the longitudinal modes do not resemble as much the modes of the graded index structure. The convergence of TE modes goes as $N^{-3}$ with the total number of modes. Finally, the accuracy of the mode obtained by re-expansion is further considered in Appendix \ref{sec:accuracy}.

\subsection{Fields excited by point source}
\label{sec:fields}
\begin{figure}[!t]
\begin{center}
\subfloat[$\real(E_z)$]{\begin{tabular}[b]{c}%
	\includegraphics{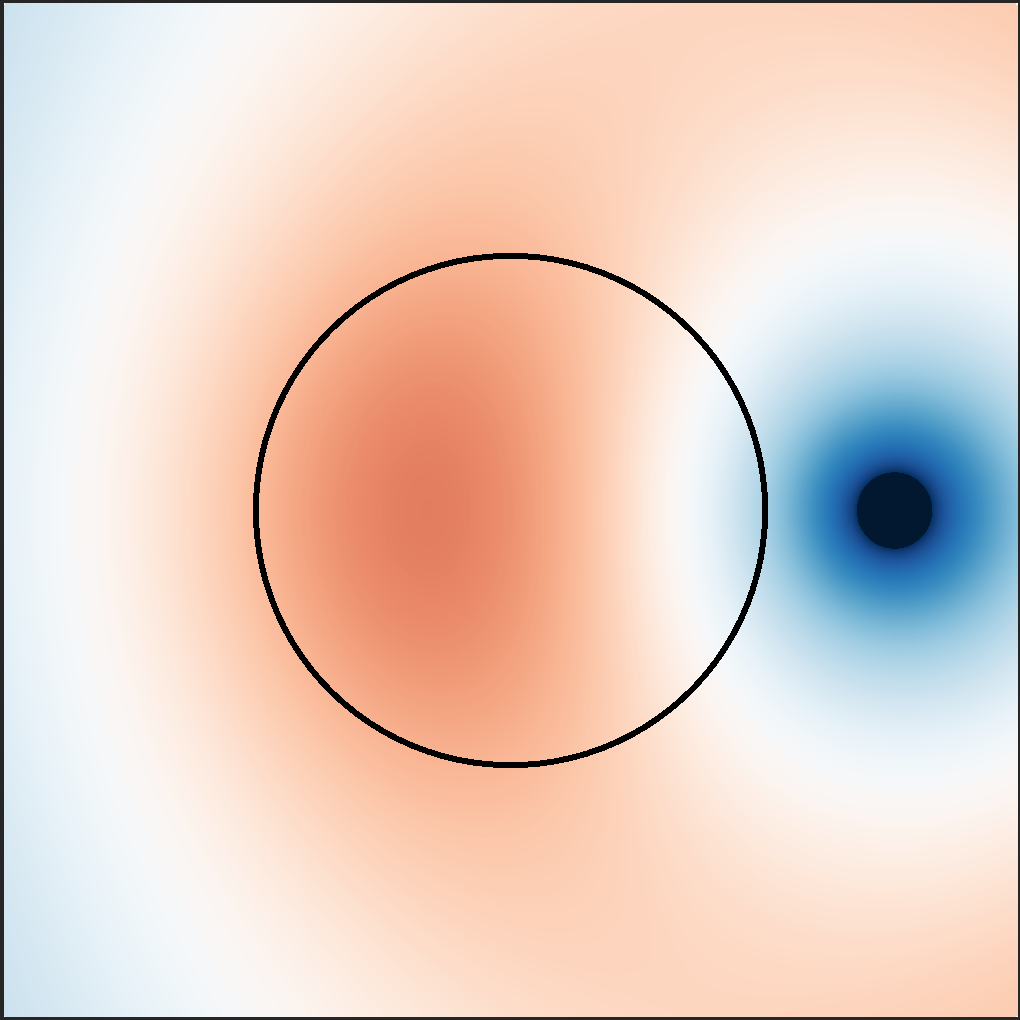}
	\end{tabular}}
\subfloat[$\imag(E_z)$]{\begin{tabular}[b]{c}%
	\includegraphics{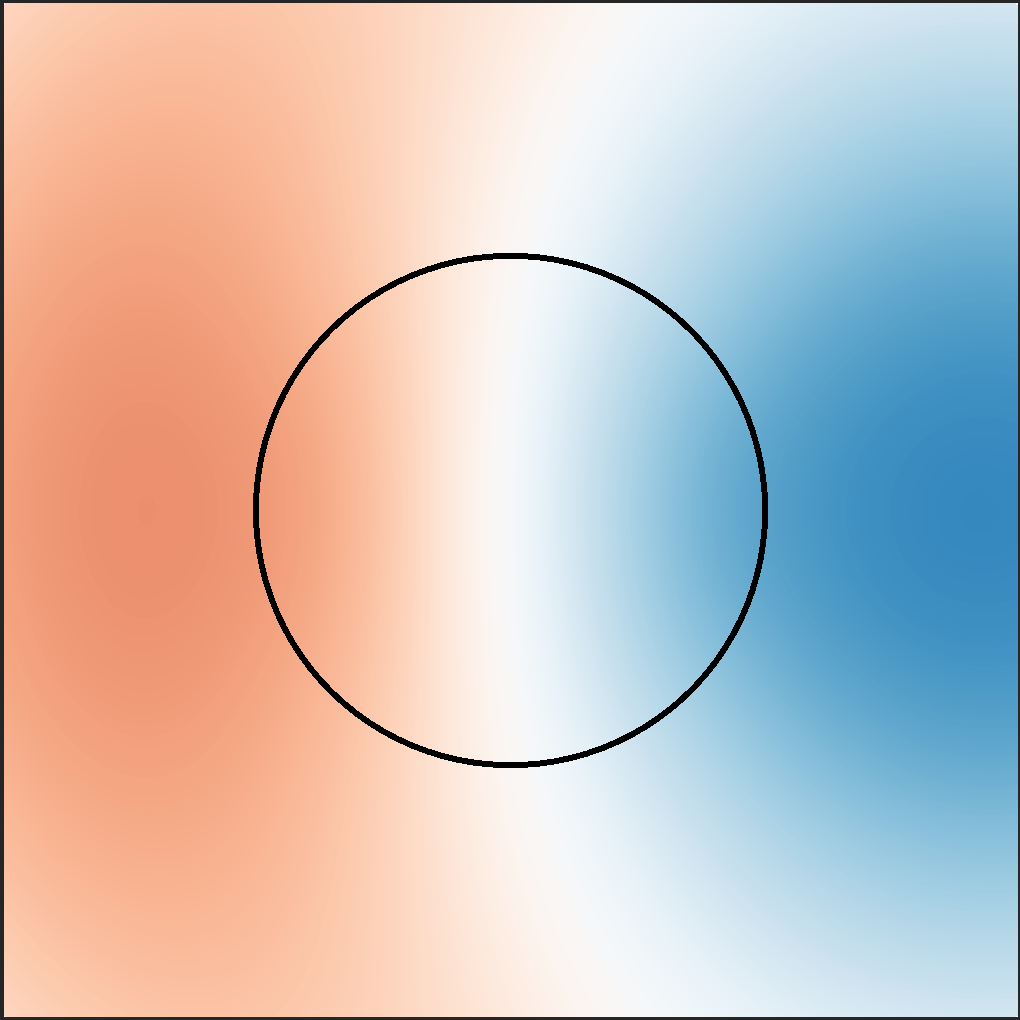}
	\end{tabular}}
\subfloat{
	\includegraphics{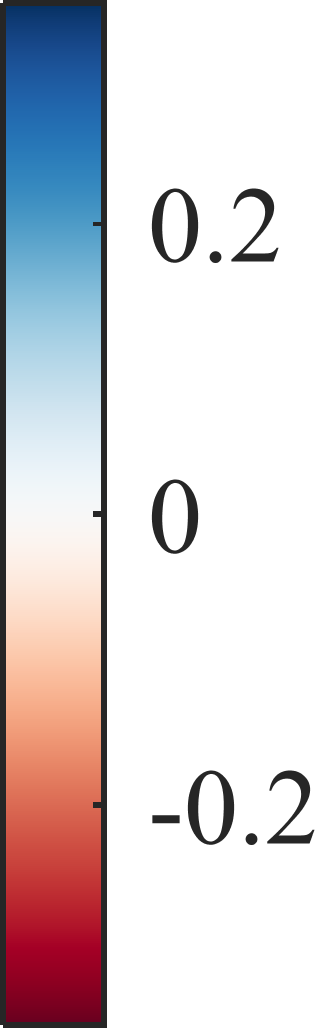}}
\caption{Shows the fields calculated using \eqref{eq:Jevform} from an oscillating out-of-plane dipole placed according to Figure \ref{fig:geometry}. Real and imaginary parts of the only non-zero component are shown.}
\label{fig:pz}
\end{center}
\end{figure}

\begin{figure}[!t]
\begin{center}
\subfloat{\begin{tabular}[b]{c}%
	\includegraphics{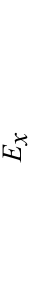}\\
	\includegraphics{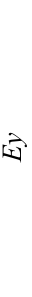}
	\end{tabular}}
\subfloat[Real part]{\begin{tabular}[b]{c}%
	\includegraphics{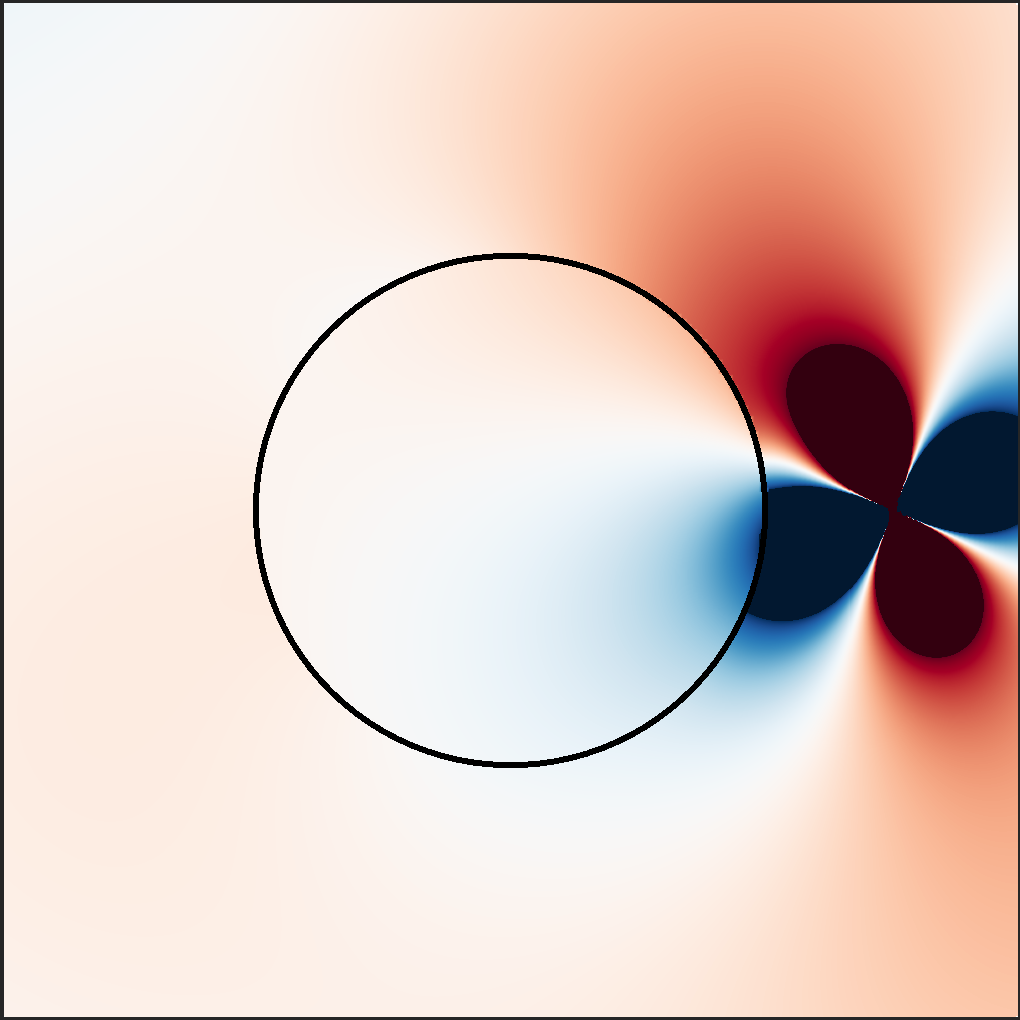}\\
	\includegraphics{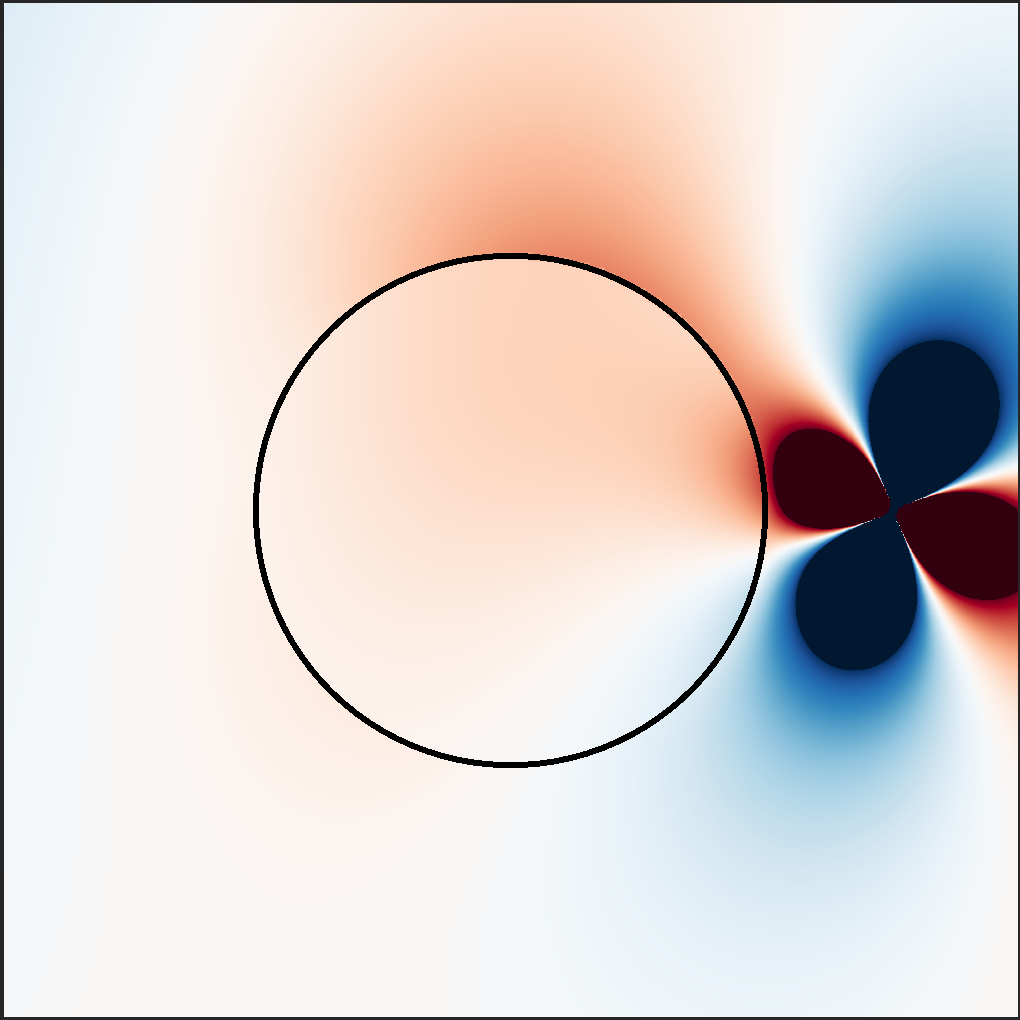}
	\end{tabular}}
\subfloat[Imaginary part]{\begin{tabular}[b]{c}%
	\includegraphics{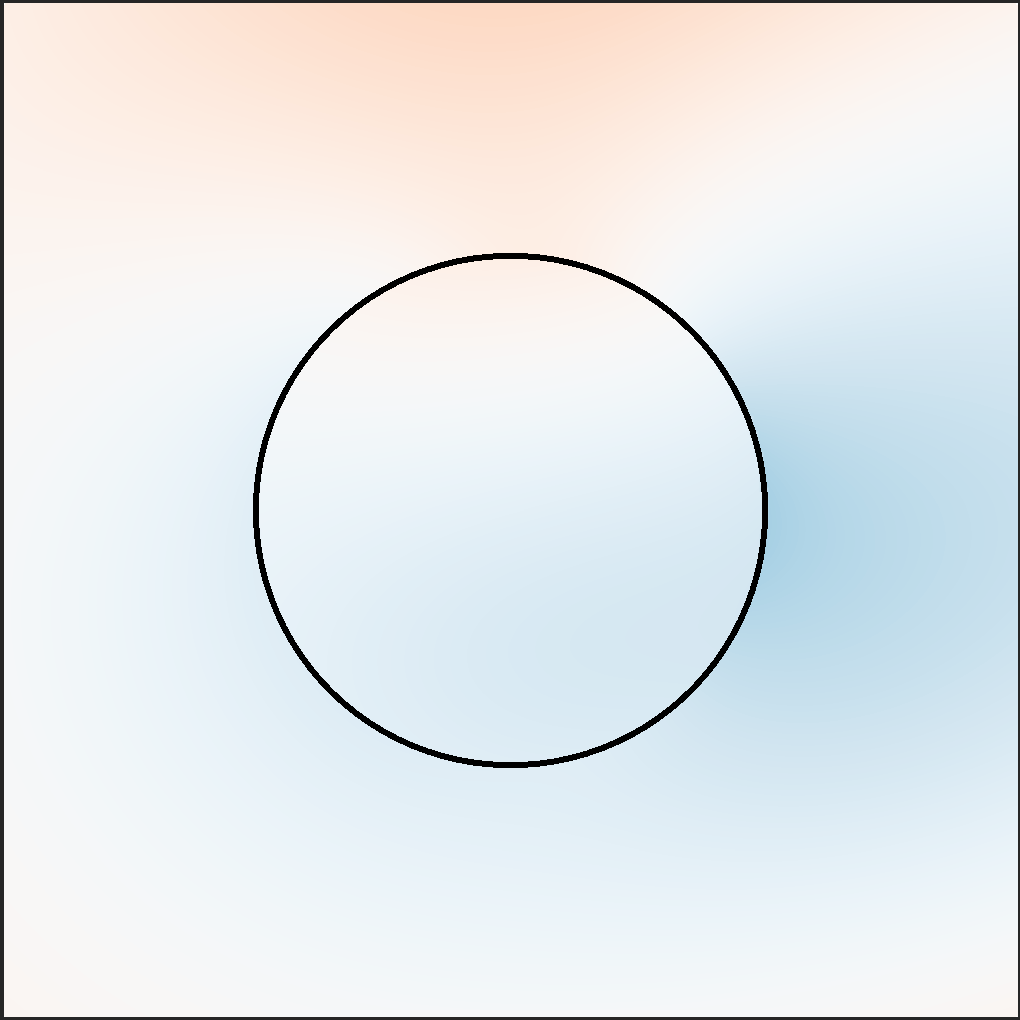}\\
	\includegraphics{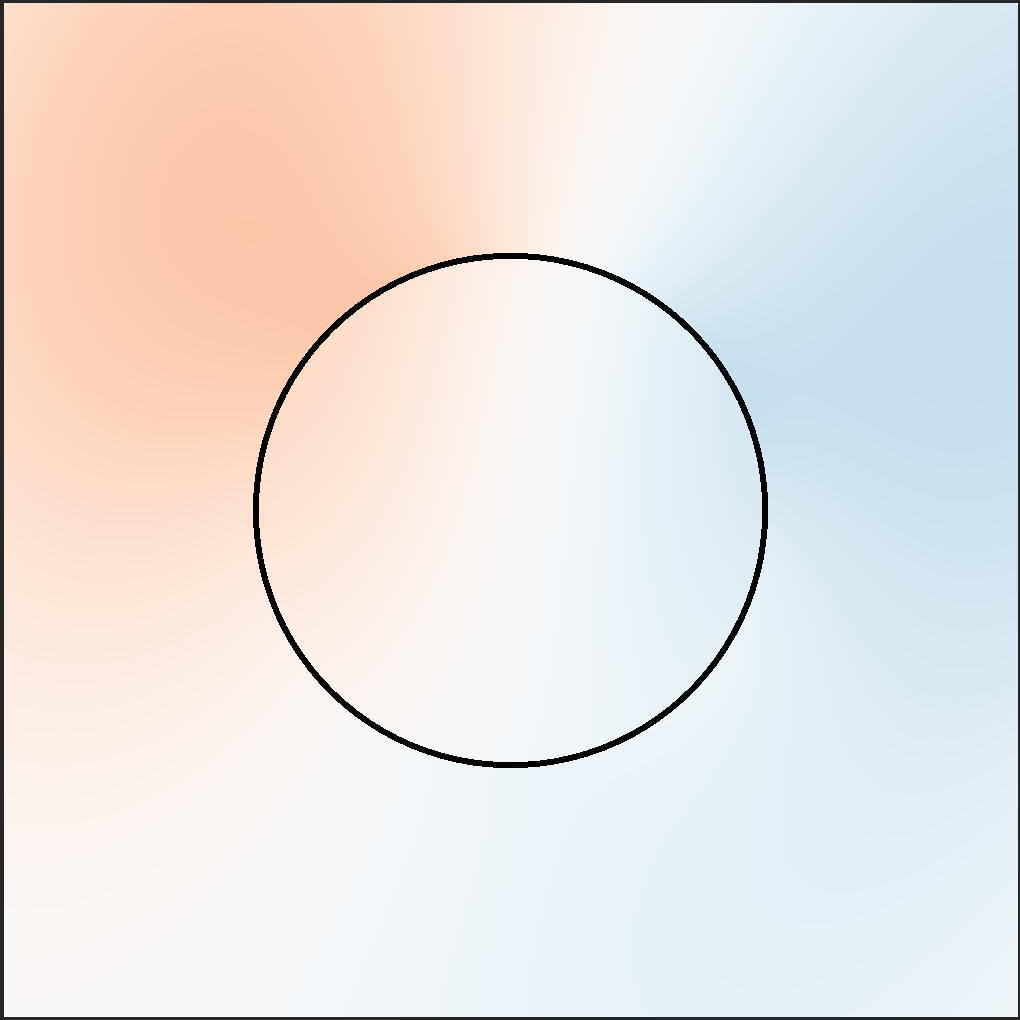}
	\end{tabular}}
\subfloat{\begin{tabular}[b]{c}%
	\includegraphics{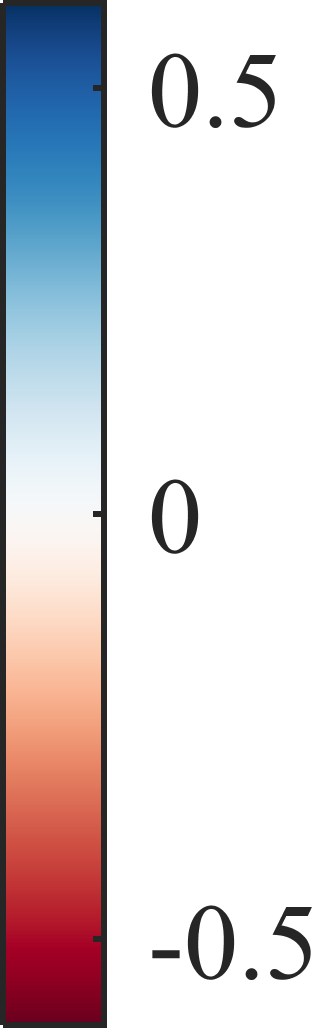}\\
	\includegraphics{pxybar.png}
	\end{tabular}}
\caption{As in Figure \ref{fig:pz}, but for an in-plane dipole moment oriented diagonally with $p_x = p_y = 1$. All non-zero field components are shown.}
\label{fig:pxy}
\end{center}
\end{figure}

Having found the eigenmodes of the graded index structure and established their accuracy, we may now use them for simulating Maxwell's equations via GENOME. We choose to excite the structure with a near field source, a 2D point dipole oscillating at a fixed frequency, corresponding to $\bv{J} = -i\omega\bv{p}_0\delta(\bv{r}-\bv{r}_0)$, where $\bv{r}_0$ denotes the location of the source. The units of dipole moment $\bv{p}_0$ in 2D differ from the more familiar 3D case, since $\bv{p}_0$ is defined per unit length in the out-of-plane direction, giving it units of Coulombs in SI units. The dipole moment can be oriented in-plane or out-of-plane, corresponding to the non-zero components of $\bv{p}_0$. This excites TE and TM modes of the graded index structure, respectively. The electromagnetic fields $\bv{E}_0$ produced by 2D point dipoles are not commonly considered in the literature, but their explicit forms are available in Ref.\ \parencite{chen2019generalizing}, for all possible orientations of $\bv{p}_0$. These $\bv{E}_0$ fields are necessary for the full solution \eqref{eq:Jevform}.

We place the source outside the resonator as shown in Figure \ref{fig:geometry}. Since the source breaks the azimuthal symmetry, we shall need modes of all azimuthal orders to represent the solution. We begin with an out-of-plane point dipole moment with $p_{0,z} = 1$ oscillating at $kB=1$. The real and imaginary parts of the total field are shown in Figure \ref{fig:pz}. The fields were generated using azimuthal orders from $-10$ to $10$, along with 50 transverse basis modes for each azimuthal order. This was sufficient to obtain the answer to high accuracy, to be demonstrated in Figure \ref{fig:maxwellerror}. All target modes found via re-expansion were inserted into GENOME \eqref{eq:greenexp}. No target modes were filtered or discarded, as no spurious modes were generated. We have reported the number of basis modes, rather than the number of target modes, since the target modes are ultimately expressed in terms of the basis modes \eqref{eq:expansion}.  The real part of the field features a logarithmic divergence at the dipole origin, characteristic of out-of-plane 2D point sources. The imaginary part remains regular, as expected. 

We now treat an in-plane point source, oriented diagonally with $p_{0,x} = p_{0,y} = 1$, also with $kB=1$. More modes are now required to achieve high accuracy, with azimuthal orders $-25$ to $25$, and 50 transverse and 50 longitudinal basis modes for each azimuthal order. The results are plotted in Figure \ref{fig:pxy}, displaying all non-zero field components. The $1/r$ divergence is present in the real part.

The convergence of the solutions is determined by the convergence of each target mode of the graded index structure against the number of basis modes, and the convergence of the total field against the number of target modes. The former was already considered in Section \ref{sec:modes}, while the latter measures the convergence properties of GENOME, which was already covered in Ref.\ \parencite{chen2019generalizing}. As such, we do not demonstrate separate convergence data here. A test of accuracy is considered in Appendix \ref{sec:accuracy}.

\section{Summary}
\label{sec:conc}
In this paper, we presented an efficient solution of electrodynamic scattering for open systems, obtaining the Green's tensor of spatially varying resonators by expanding in terms of its generalized normal modes. The formulation for GENOME was generalized to treat such profiles, culminating in \eqref{eq:E0solus} and \eqref{eq:greenexp}. These expressions are almost unchanged compared to the formulation for uniform inclusions, requiring only a different set of modes, and thus retain all previously discussed advantages of GENOME.\autocite{chen2019generalizing} We then presented an efficient and reliable method of obtaining these modes, by re-expanding into a set of basis modes, which are the solutions of a simpler open system. This procedure requires only the solution of a linear matrix eigenvalue problem \eqref{eq:perteig}, populated by the overlap integrals between basis modes \eqref{eq:Vdef}. Although the re-expansion method requires a set of longitudinal modes, these are simple to obtain, via \eqref{eq:potendef} and \eqref{eq:Lhelmholtz}.

We demonstrated the implementation of the re-expansion method, successfully obtaining the modes of a graded index fiber (Figure \ref{fig:geometry}). Rapid convergence of both TM and TE modes was observed, as shown in Figures \ref{fig:conv} and \ref{fig:TEconv}, demonstrating that the matrix eigenvalue problem \eqref{eq:perteig} remains small. Next, we employ these modes via our generalization of GENOME \eqref{eq:greenexp} to solve a scattering problem of this graded index structure excited by a near-field source. The accuracy of the solution was confirmed, again demonstrating one of the key strengths of GENOME, in that the solution is correct over all space, even for sources and detectors exterior to the resonator. 

\section*{Acknowledgments}
P.\ Y.\ C.\ and Y.\ S.\ were partially supported by Israel Science Foundation (ISF) Grant No.\ 889/16.

\appendix
\section{Orthogonality of basis modes}
\label{sec:ortho}
Orthogonality is an important property that permits projection and ultimately the derivation of Section \ref{sec:perturbation} to proceed. We shall demonstrate that the basis modes satisfy the orthogonality relation \eqref{eq:ortho}, confirming the simple form of the adjoint modes $\bv{E}^\dagger_m$ in the process. This leads to a simple projection operator \eqref{eq:project} that decomposes any given field into a sum over the basis modes, which we also derive.

The adjoint modes can be obtained by taking the transpose of eigenvalue equation \eqref{eq:eigenint}, and invoking reciprocity, the symmetry of the Green's tensor both under transposition and the interchange of $\bv{r}$ and $\bv{r}'$, $\tensor{G}_0^\intercal(|\bv{r}-\bv{r}'|) = \tensor{G}_0(|\bv{r}'-\bv{r}|)$,
\begin{equation}
s_m \bv{E}^\dagger_m(\bv{r}) = k^2 \epsilon_b \int \bv{E}^\dagger_m(\bv{r}') \epsilon_C(\bv{r}') \tensor{G}_0 (|\bv{r}' - \bv{r}|)  \, d\bv{r}'.
\label{eq:eigenadj}
\end{equation}
We have identified $\bv{E}^\dagger(\bv{r}) = \bv{E}^\intercal(\bv{r})$ as the adjoint modes. The proof that such modes are the correct adjoint and possess the necessary properties for orthogonality \eqref{eq:ortho} is similar to other such proofs, proceeding from a construction that involves the operator that generates the eigenmodes,
\begin{equation}
\begin{aligned}
&k^2\epsilon_b\iint \bv{E}_n^\dagger(\bv{r}) \epsilon_C(\bv{r}) \tensor{G}_0(|\bv{r}-\bv{r}'|) \epsilon_C(\bv{r}') \bv{E}_m(\bv{r}')\, d\bv{r} d\bv{r}'\\
&= s_m \int \bv{E}_n^\dagger(\bv{r}) \epsilon_C(\bv{r}) \bv{E}_m(\bv{r})\, d\bv{r}.
\end{aligned}
\end{equation}
The result was obtained using the defining eigenvalue equation \eqref{eq:eigenint}. Using instead the adjoint equation \eqref{eq:eigenadj}, the construction evaluates to
\begin{equation}
s_n \int \bv{E}_n^\dagger(\bv{r}') \epsilon_C(\bv{r}') \bv{E}_m(\bv{r}')\, d\bv{r}'.
\end{equation}
Combining these results gives
\begin{gather}
(s_n - s_m) \int \bv{E}_n^\dagger(\bv{r}) \epsilon_C(\bv{r}) \bv{E}_m(\bv{r})\, d\bv{r} = 0,
\label{eq:orthoproof}
\end{gather}
stating that eigenmodes belonging to different eigenvalues are orthogonal with weight function $\epsilon_C(\bv{r})$, and confirming that \eqref{eq:adjoint} are the adjoint modes. Finally, upon normalizing each mode, we obtain the desired relation \eqref{eq:ortho}.

The primary purpose of orthogonality is to provide a simple projection procedure \eqref{eq:project}. This decomposes an arbitrary field $\bv{E}(\bv{r})$ in the interior of the geometry, where the basis is complete, as
\begin{equation}
\bv{E}(\bv{r}) = \sum_m \bv{E}_m(\bv{r}) \int \bv{E}^\dagger_m(\bv{r'}) \epsilon_C(\bv{r'}) \bv{E}(\bv{r'})\, d\bv{r'}.
\label{eq:decompose}
\end{equation}
The derivation of this property from orthogonality follows from completeness,
\begin{equation}
\bv{E}(\bv{r}) = \sum_m c_m \bv{E}_m(\bv{r}).
\end{equation}
The coefficients $c_m$ are then found by projecting onto $\bv{E}^\dagger_n(\bv{r})$ and using the orthogonality \eqref{eq:ortho} just derived,
\begin{equation}
\int \bv{E}^\dagger_n(\bv{r})\epsilon_C(\bv{r})\bv{E}(\bv{r})\, d\bv{r} = \sum_m c_m \int \bv{E}^\dagger_n(\bv{r}) \epsilon_C(\bv{r}) \bv{E}_m(\bv{r})\, d\bv{r} = \sum_m c_m \delta_{nm},
\end{equation}
leading to \eqref{eq:decompose} and \eqref{eq:project}.

\section{Transverse basis modes}
\label{sec:transverse}
The longitudinal modes of Section \ref{sec:basis} are irrotational by definition, so ordinary transverse basis modes are required for a complete basis. Transverse basis modes are also entirely responsible for the far-field features of the target modes, as longitudinal modes do not carry energy away from the resonator. We briefly overview the basis modes of a uniform 2D cylinder (see Figure \ref{fig:profile}). These well known modal fields can be obtained analytically, giving also a transcendental equation for the eigenvalues. Very similar procedures apply in 1D and 3D.

Applying separation of variables to the Lippmann-Schwinger equation \eqref{eq:eigendiff} yields the functional form of the eigenmodes. Since the system is invariant in one dimension, two distinct polarizations exist: transverse electric and transverse magnetic. For the TM modes, the fields are given by
\begin{equation}
\begin{aligned}
\tilde{E}_z &= H^{(1)}_\tau(\sqrt{\epsilon_b}kr) e^{i\tau\vartheta}, & &\bv{r} \in \textrm{exterior}\\
\tilde{E}_z &= J_\tau(\sqrt{\epsilon_{\tau\tau'}}kr) e^{i\tau\vartheta}, & &\bv{r} \in \textrm{interior}.
\end{aligned}
\label{eq:TMmode}
\end{equation}
The eigenpermittivity $\sqrt{\epsilon_{\tau\tau'}}$ is yet to be determined. For brevity, we display the complex exponential forms, but the sine and cosine forms can be used in an implementation. As with \eqref{eq:FBmode}, there are two integer subscripts.

The TE modes have similar form, but it is simpler to specify the magnetic fields
\begin{equation}
\begin{aligned}
H_z &= H_\tau(\sqrt{\epsilon_b}kr) e^{i\tau\vartheta}, & &\bv{r} \in \textrm{exterior}\\
H_z &= J_\tau(\sqrt{\epsilon_{\tau\tau'}}kr) e^{i\tau\vartheta}, & &\bv{r} \in \textrm{interior},
\end{aligned}
\label{eq:TEmode}
\end{equation}
and use the Maxwell curl equation $\bv{E} = \frac{i}{k\epsilon} \nabla\times(Z_0\bv{H})$ to obtain the electric fields when performing the integrals of \eqref{eq:Vdef}.

The eigenpermittivity can be determined by applying Maxwell's boundary conditions at the boundary between the interior and exterior, defined at radius $B$. This leads to a form of the well-known dispersion relation for a step-index fiber with zero propagation wave number along the fiber,
\begin{equation}
\left(\frac{1}{\sqrt{\epsilon_{\tau\tau'}} kB}\frac{J'_\tau(\sqrt{\epsilon_{\tau\tau'}} kB)}{J_\tau(\sqrt{\epsilon_{\tau\tau'}} kB)} - \frac{1}{\sqrt{\epsilon_b} kB}\frac{H'_\tau(\sqrt{\epsilon_b} kB)}{H_\tau(\sqrt{\epsilon_b} kB)}\right) \left(\frac{\sqrt{\epsilon_{\tau\tau'}}}{kB}\frac{J'_\tau(\sqrt{\epsilon_{\tau\tau'}} kB)}{J_\tau(\sqrt{\epsilon_{\tau\tau'}} kB)} - \frac{\sqrt{\epsilon_b}}{kB}\frac{H'_\tau(\sqrt{\epsilon_b} kB)}{H_\tau(\sqrt{\epsilon_b} kB)}\right) = 0.
\label{eq:disprel}
\end{equation}
Eigenvalues associated with different azimuthal orders $\tau$ originate from different orders of \eqref{eq:disprel}, while different radial orders $\tau'$ are different roots of the same azimuthal order.

\section{Accuracy tests}
\label{sec:accuracy}
\begin{figure}[!t]
\begin{center}
\subfloat{\begin{tabular}[b]{c}%
        \begin{tabular}{l}%
	\includegraphics{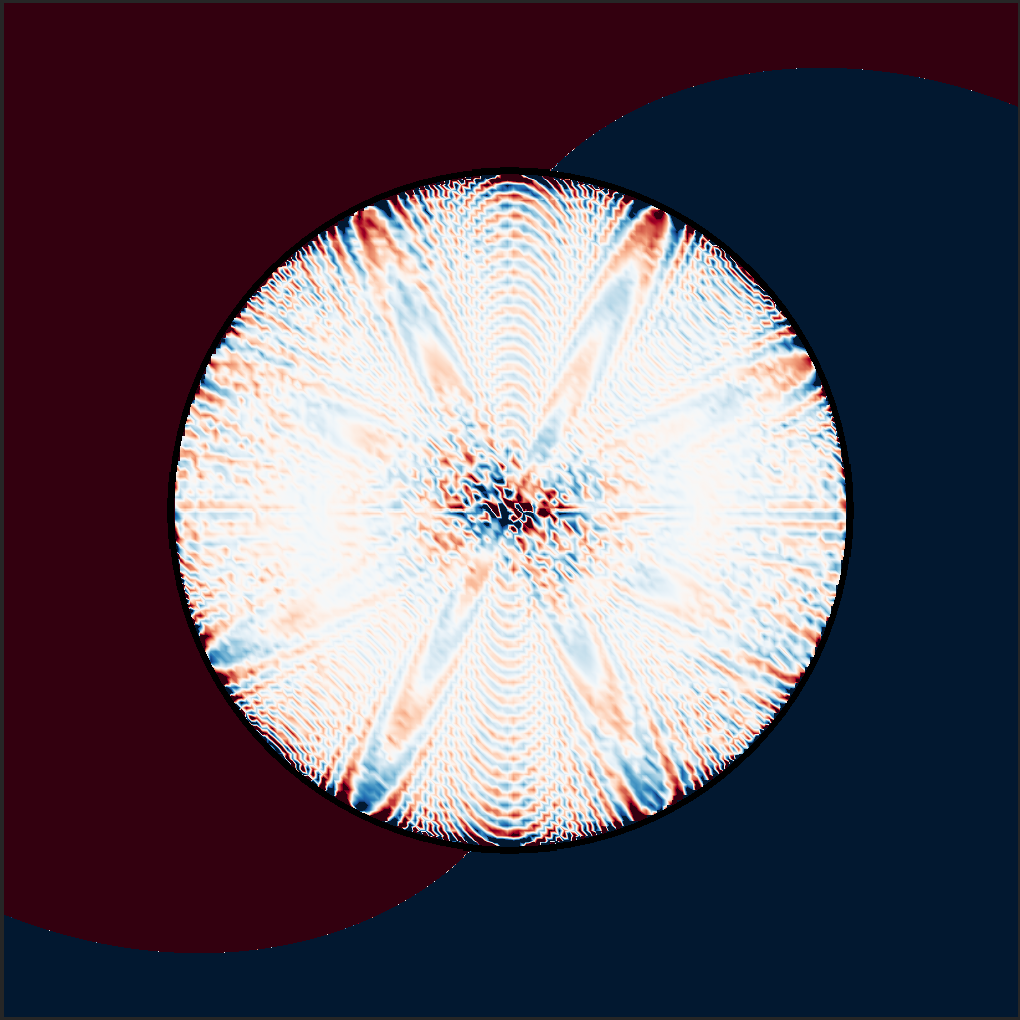}
	\end{tabular}\end{tabular}}
\subfloat{\begin{tabular}[b]{c}%
        \begin{tabular}{l}%
	\includegraphics{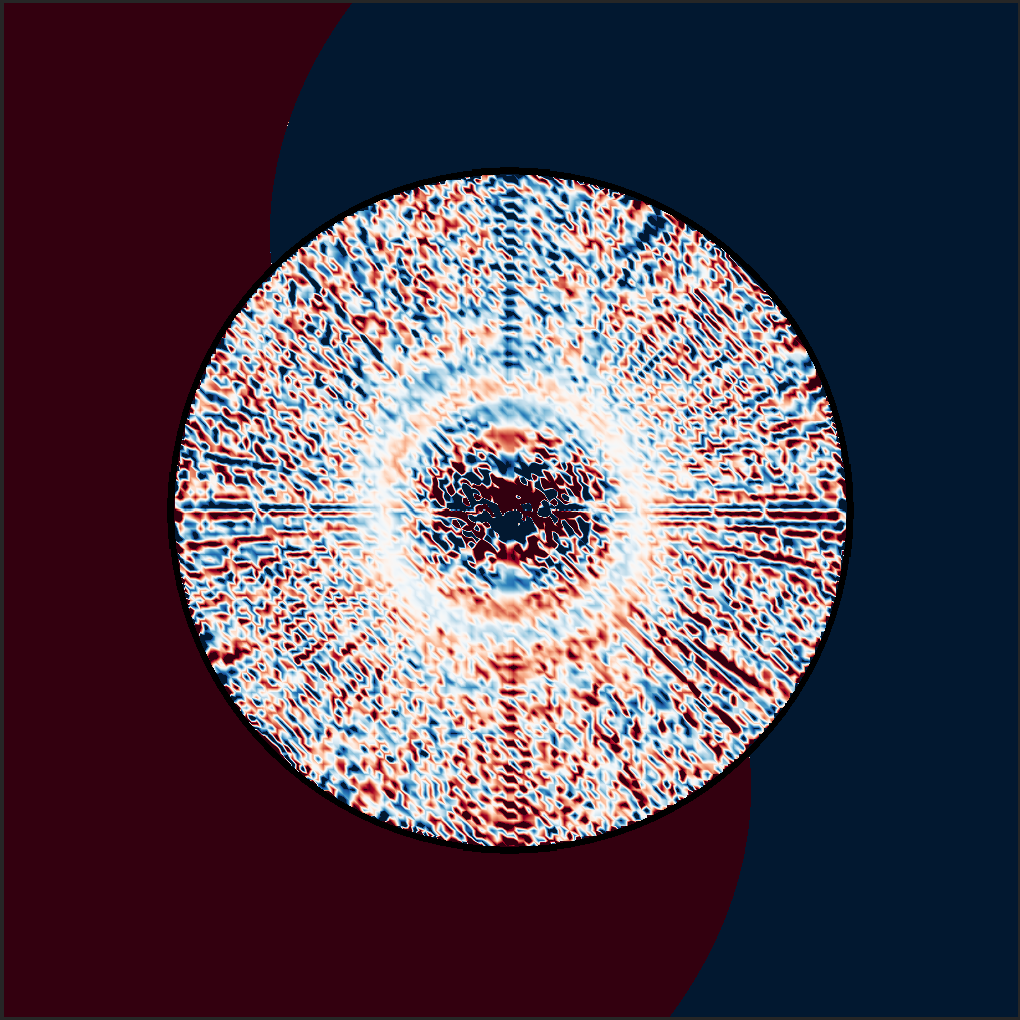}
	\end{tabular}\end{tabular}}
\subfloat{\begin{tabular}[b]{c}%
        \begin{tabular}{l}%
	\includegraphics{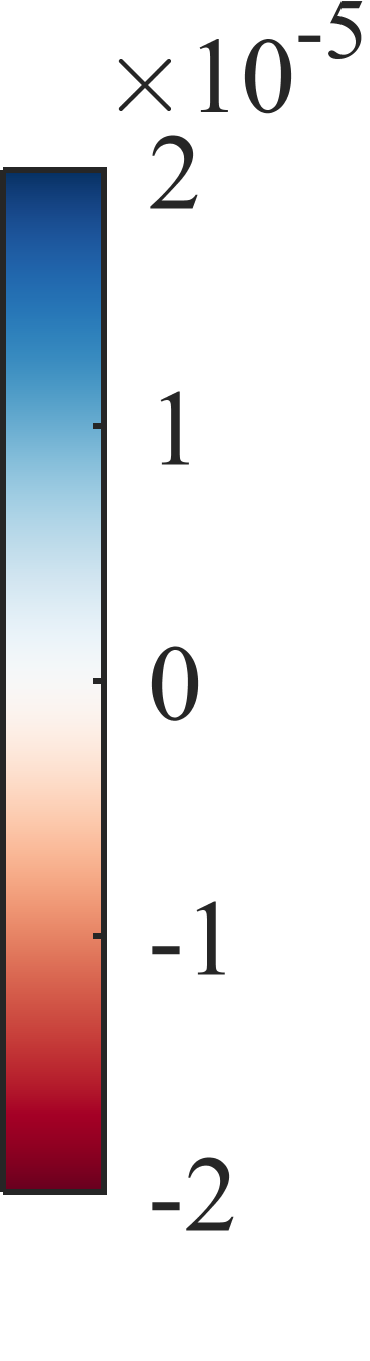}
	\end{tabular}\end{tabular}}
\caption{Difference between the two sides of equation \eqref{eq:eigenexpanded} for the two TE modes of Figure \ref{fig:TEmodes}. The test is performed on the $\real(E_z)$ component. The test applies only to the interior region, so the parts of the plots for the background region should be disregarded.}
\label{fig:eqnconfirm}
\end{center}
\end{figure}

\begin{figure}[!t]
\begin{center}
\subfloat[$\real(E_z)$]{\begin{tabular}[b]{c}%
        \begin{tabular}{l}%
	\includegraphics{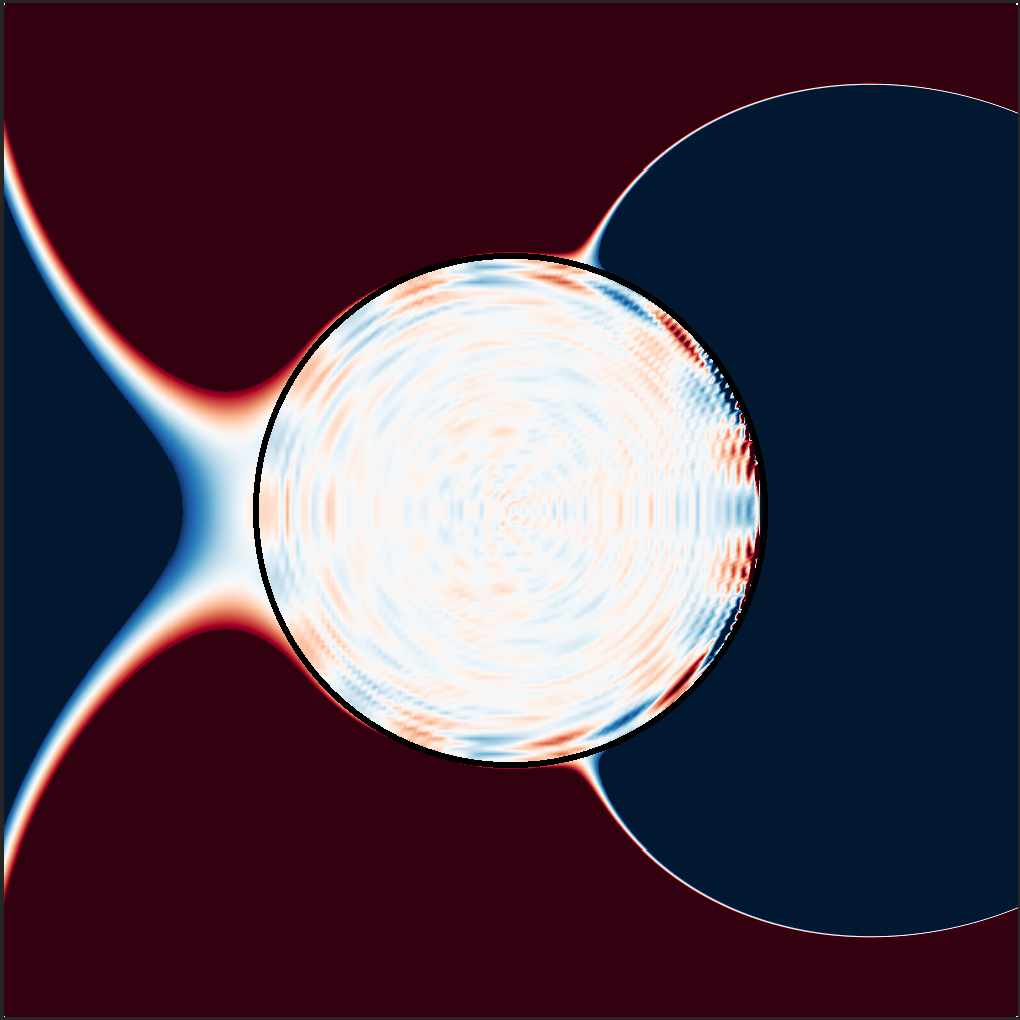}
	\end{tabular}\end{tabular}}
\subfloat[$\real(E_x)$]{\begin{tabular}[b]{c}%
        \begin{tabular}{l}%
	\includegraphics{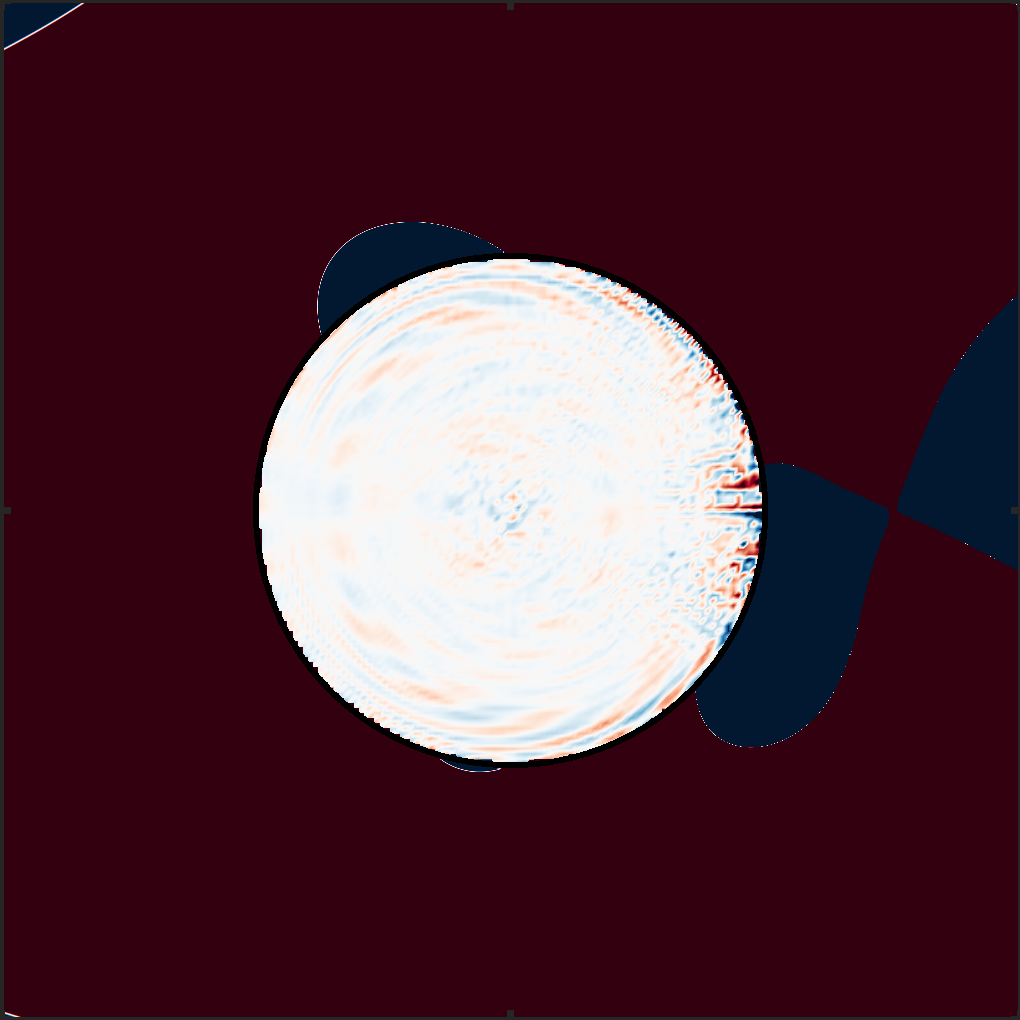}
	\end{tabular}\end{tabular}}
\subfloat{\begin{tabular}[b]{c}%
        \begin{tabular}{l}%
	\includegraphics{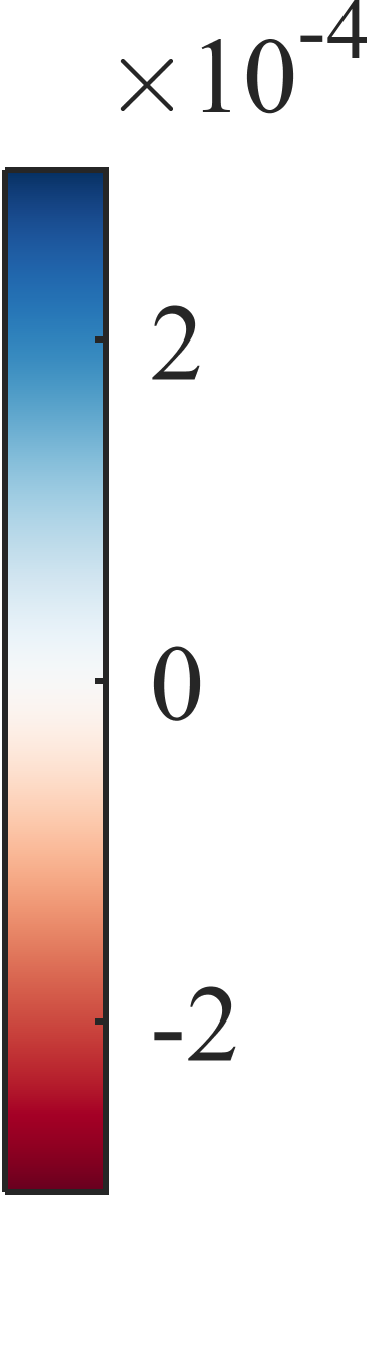}
	\end{tabular}\end{tabular}}
\caption{LHS of \eqref{eq:maxcond}. For (a), the test is performed on the $\real(E_z)$ component of the solution in Figure \ref{fig:pz}, while for (b), the test is performed on the $\real(E_x)$ component of Figure \ref{fig:pxy}. The same number of modes were used here as in those figures. The test applies only to the interior region, so the parts of the plots for the background region should be disregarded.}
\label{fig:maxwellerror}
\end{center}
\end{figure}

We have demonstrated the convergence properties of the re-expansion method in Sections \ref{sec:modes} and \ref{sec:fields}, but this of course is not sufficient to demonstrate its accuracy. Since the method converges faster than methods that rely on spatial discretization for example, it is preferable to perform self-consistency checks than to rely on comparisons to other methods. We may certify the modes to be accurate if they satisfy the governing differential equation \eqref{eq:eigendiff} over all space. Within the inclusion interior, confirming that \eqref{eq:eigendiff} is satisfied is best achieved by numerically confirming that the two sides of \eqref{eq:eigenexpanded} are equal. We perform this test for the two fundamental TE modes normalized according to \eqref{eq:ortho}, plotting in Figure \ref{fig:eqnconfirm} the difference between the two sides of \eqref{eq:eigenexpanded}. The test is confirmed to 5 or more digits of precision within the inclusion interior for both modes, thus confirming their accuracy. The modes are possibly accurate to more digits, as this test only sets a lower bound on their accuracy. The test is valid only within the interior, since \eqref{eq:eigendiff} is valid only in the interior. So only the interior regions of Figure \ref{fig:eqnconfirm} are pertinent, and the exterior regions merely feature numerical noise that should be disregarded. Instead, Maxwell's equations are satisfied by construction by the expansion \eqref{eq:expansion} in the infinite background. This holds because each basis mode satisfies Maxwell's equations here, as does any linear combination of basis modes.

Next, we confirm the accuracy of the fields in Figures \ref{fig:pz} and \ref{fig:pxy} calculated using GENOME. We again seek to verify that Maxwell's equations are satisfied over all space. It is also satisfied by construction in the infinite exterior, including the divergence at the source. For the interior, we could perform the $\nabla\times\nabla\times$ operator numerically, but to avoid numerical errors, we may derive an analytic condition by inserting the GENOME expansion \eqref{eq:Jevform} into Maxwell's equations \eqref{eq:maxwell}, yielding
\begin{equation}
\nabla\times(\nabla\times\bv{E}_0) - k^2\epsilon(\bv{r})\bv{E}_0 + \sum_m w_m [\nabla\times\nabla\times \bv{E}_m - k^2\epsilon(\bv{r})\bv{E}_m] = ikZ_0\bv{J},
\end{equation}
using $w_m$ to denote all the factors associated with each mode in the sum of \eqref{eq:Jevform}. In our case, $\bv{J}$ is zero, since we are inside the resonator. The $\nabla\times\nabla\times\bv{E}_m$ can be simplified by the eigenvalue equation \eqref{eq:eigendiff}, while $\bv{E}_0$ satisfies
\begin{equation}
\nabla\times(\nabla\times\bv{E}_0) - k^2\epsilon_b\bv{E}_0 = ikZ_0\bv{J}.
\end{equation}
After some algebra, we obtain the condition
\begin{equation}
\bv{E}_0 + \sum_m w_m \left(1-\frac{1}{s_m}\right) \bv{E}_m = 0,
\label{eq:maxcond}
\end{equation}
valid for the interior. We plot the LHS of \eqref{eq:maxcond} in Figure \ref{fig:maxwellerror}, for both the out-of-plane example of Figure \ref{fig:pz} and the in-plane example of Figure \ref{fig:pxy}. As can be seen, \eqref{eq:maxcond} is satisfied to more than 4 or 5 digits of precision. 

\printbibliography
\typeout{get arXiv to do 4 passes: Label(s) may have changed. Rerun}
\end{document}